\documentclass[12pt]{article}
\usepackage{subfig}
\usepackage{youngtab}
\AtBeginDocument{
	\addtocontents{toc}{\normalsize}
}
\pdfoutput=1

\def\denom{{\mathtt D}}

\usepackage{amsmath,amssymb,amsfonts,amsthm,amscd,mathrsfs}
\usepackage{esvect}
\usepackage{xcolor}
\definecolor{darkblue}{rgb}{0.1,0.1,.7}
\usepackage[]{version}
\usepackage[]{graphicx}
\usepackage[]{latexsym}
\usepackage{geometry}
\usepackage[all,cmtip]{xy}
\usepackage[margin=10pt,font=small,labelfont=bf]{caption}
\usepackage{ifthen}
\usepackage{tikz}
\usepackage{array,setspace,mathrsfs,amsfonts,yfonts,dsfont,bbm,colonequals}
\usepackage{dsfont}
\usepackage{cite}
\usepackage{xspace}
\usepackage{empheq}
\usepackage{extarrows}
\usepackage{braket}
\usepackage{mathtools}
\usepackage[numbers,sort&compress]{natbib}
\usepackage{cite}
\usepackage{hyperref}
\usepackage[utf8]{inputenc}
\usepackage{setspace}
\usepackage{float, graphicx}
\usepackage{dsfont,hyperref}
\numberwithin{equation}{section}
\def\ts{{\mathtt s}}
\def\tv{{\mathtt v}}
\def\tt{{\mathtt t}}
\def\CM{{\mathcal M}}
\def\tg{{\mathtt g}}

\newcommand{\ve}{\varepsilon}

\newcommand{\g}{\gamma}

\newcommand{\be}{\begin{equation}}
\newcommand{\ee}{\end{equation}}
\newcommand{\bea}{\begin{eqnarray}}
\newcommand{\eea}{\end{eqnarray}}
\newcommand{\ba}{\begin{equation}\begin{aligned}}
\newcommand{\ea}{\end{aligned}\end{equation}}

\newcommand{\R}{\mathbb{R}}

\newcommand{\Z}{\mathbb{Z}}
\newcommand{\tyng}{\tiny\yng}

\def\g{\gamma}

\def\s{\sigma}

\def\a{\alpha}

\def\b{\beta}
\def\d{\delta}

\newcommand{\syng}[1]{\scalebox{0.2}{\yng(#1)}}

\def\ve{{\varepsilon}}
\def\s{{\sigma}}
\def\g{{\gamma}}

\def\a{{\alpha}}
\def\b{{\beta}}
\def\c{{\gamma}}
\def\d{{\delta}}

\def\CM{{\mathcal M}}

\def\CS{{\mathcal S}}

\def\CV{{\mathcal V}}

\newcommand{\bd}[1]{\begin{fmffile}{#1}\begin{fmfgraph*}}
		\newcommand{\ed}{\end{fmfgraph*}\end{fmffile}}

\begin{document}

\title{\bf Counting parity-violating local S-matrices}
\date{}
\author{Subham Dutta Chowdhury$^{a}$\footnote{subham@uchicago.edu}\\
	\it\\
	\it $^{a}$ Enrico Fermi Institute $\&$ Kadanoff Center for Theoretical Physics,\\
	\it University of Chicago, Chicago, IL 60637, USA\\
}
\maketitle
\vskip 2cm
\abstract{Four point tree-level local S-matrices form a module over ring of polynomials of mandelstam invariants s, t and u. The module of local analytic S-matrices can be encoded in terms of a partition function which is enumerated using plethystic techniques. In this paper, we enumerate the plethystic contribution to local four point photon, graviton and gluon multi-particle partition functions that encode parity violating $2 \rightarrow 2$ scattering. We generalise the counting problem solved in \cite{ Chowdhury:2019kaq, Chowdhury:2020ddc} to project out parity violating sectors, a subtle task in even dimensions \cite{Henning:2017fpj}. We explicitly enumerate the parity odd contributions to the multi-letter partition function for gauge fields, gravitons and gluons and evaluate the resulting parity violating partition functions in $D=4,6$. We also perform a large $D$ analysis to show that parity violating local interactions do not contribute to four particle scattering in higher dimensions ($D\geq 8$). Our computations and observations for photons, gravitons and gluons agree with the transformation properties of these S-matrices previously conjectured in \cite{Chowdhury:2019kaq, Chowdhury:2020ddc}}

\tableofcontents
\section{Introduction}

 Let us consider  scattering of four massless particles, in $D$ space-time dimensions, obeying the following momentum conservation and mass shell conditions 
\be
(p^{(i)})^2=0,\qquad \qquad \sum_i\,  p_\mu^{(i)}=0.
\ee
where $p_\mu^{(i)}$ denotes the momenta of the $i$th particle. We define the mandelstam invariants for $2 \rightarrow 2$ scattering as follows,

\begin{equation}\label{stu} \begin{split}
s&=-(p_1+p_2)^2=-2 p_1.p_2, \qquad t=-(p_1+p_3)^2=-2 p_1.p_3,\qquad u=-(p_1+p_4)^2=-2 p_1.p_4
\end{split}
\end{equation}

We are interested in the space of S-matrices which are analytic functions of the mandelstam invariants $s,t$ (since $s+t+u=0$ for massless scattering). The reader can assume that these will be generated by contact term like quartic lagrangians of arbitrarily high derivative orders corresponding to different species of particles. The S-matrix is also characterised by the polarisation data (except for scalars, of course) and colour labels that specifies the internal spin degree of freedom and the colour degrees of the scattering particles respectively. The internal spin and colour degrees of freedom impose non-trivial constraints on the Lorentz-invariant four particle S-matrix. In \cite{ Chowdhury:2019kaq, Chowdhury:2020ddc}, analytic S-matrices of photons, gravitons and gluons, in arbitrary dimensions, were characterised and classified as $S_4$ invariant Lorentz invariant analytic polynomials of momenta and polarisations, where $S_4$ is the permutation group of four objects. The space of S-matrices inherit a module structure over ring of polynomials of mandelstam invariants and a systematic explicit classification of the module generators and their derivative ``descendants'' was also done. For high space-time dimensions the generators are parity invariant, finite in number and the number stabilizes beyond $D\geq 8$. For $D\leq 8$, additional parity violating  generators appeared along with reduction of the parity invariant generators. In \cite{Henning:2017fpj, Chowdhury:2019kaq}, it was shown that the space of local S-matrices are in one to one correspondence with the space of local lagrangians modulo equations of motion and total derivatives. Local lagrangians at the lowest derivative order can be thought of as the generator for the module. The descendants of the module generator then corresponds to higher derivative lagrangians constructed by sprinkling contracted derivatives on the lowest derivative lagrangian. The space of local lagrangians can also be enumerated using plethystic techniques which served as a cross-check of the module classification\footnote{These techniques are also referred to in literature as Hilbert series and have been used in several related contexts in recent times \cite{Graf:2022rco,Graf:2020yxt,Henning:2015alf,Kobach:2017xkw, Melia:2020pzd}. In the context of CFTs, similar kind of techniques have been used to determine independent tensor structures of four point functions which are in one-to-one correspondence with flat space scattering \cite{Kravchuk:2016qvl}}. However, in the lower even dimensions, the partition function contributions of the parity violating modules could not be separated from the parity invariant ones and the transformation properties of the parity violating modules were guessed from the explicit module constructions and differences from the higher dimensional partition function since parity violating modules are absent in higher dimensions.
In odd dimensions, the parity violating lagrangians are odd number in derivatives and their contribution to the plethystic partition function could be easily distinguished. The aim of this note is to fill in this gap by explicitly evaluating the parity violating contribution to the multi-letter particle functions for photons, gravitons and gluons relevant for $2 \rightarrow 2$ parity violating scattering in even dimensions. We begin by reviewing some generalities for three different types of scattering we consider.

\subsubsection*{Scalars}
Four point local Lorentz-invariant S-matrices of scalars are given by polynomials of the mandelstam invariants $s,t,u$ obeying the condition $s+t+u =0$. Let us denote them by $\mathcal{S}^{\ts} (p^{(i)})$, where $i$ denotes the particle label. We can also consider scalar particles charged under some internal symmetry group $G$. For scalar charged under the adjoint representation, the scalar degree of freedom are labelled by the adjoint generators $\tau^a$. Consequently the S matrix must be also be a G-singlet, i.e singlet under the group rotation 
\be 
\tau^{(i),a}\to R^a_{\, b} \, \tau^{(i),b} 
\ee  
The resulting S-matrix will be denoted by $\mathcal{S}^{\ts, G} (\tau^{(i),a}, p_\mu^{(i)})$. 
\subsection*{Photons}
Massless photons are characterised by the $D$-dimensional polarisation tensor $\epsilon^{(i)}_\mu$. We choose to work in the Lorentz gauge 
\be
\epsilon^{(i)}.p^{(i)}=0
\ee 
Invariance under residual gauge transformations imply that the S matrix $\mathcal{S}^{\tv} (p^{(i)}, \epsilon^{(i)})$ must be invariant under the linearised gauge transformations, 
\be\label{gaugeph}
\epsilon_{\mu}^{(i)} \to \epsilon_{\mu}^{(i)} +p_\mu^{(i)} \xi^{(i)},\qquad \forall i \in(1,4)
\ee
where $\xi^{(i)}$ is a scalar function of $p^{(i)}$. For the scattering of four massless photons, $\mathcal{S}^{\tv} (p^{(i)},\epsilon^{(i)})$ is also linearly homogeneous in each of the polarisation vectors $\epsilon^{(i)}_\mu$. 
 
\subsection*{Gravitons}
 The spin degrees of freedom of the graviton is parametrized by a symmetric traceless tensor $h_{\mu \nu}$ obeying the Lorentz gauge condition $p^\mu h_{\mu \nu}=0$. No generality is lost in restricting to the case 
 
 \be 
 h^{(i)}_{\mu \nu}= \epsilon^{(i)}_\mu \epsilon^{(i)}_\nu,\qquad p^{(i)} \cdot \epsilon^{(i)}= \epsilon^{(i)}\cdot \epsilon^{(i)} =0    
 \ee 
where the first part of the second equality is the Lorentz gauge condition while the second part implements the tracelessness condition. Thus the graviton S-matrix $\mathcal{S}^{\tt} (p^{(i)}, \epsilon^{(i)}\epsilon^{(i)})$ is a Lorentz invariant function of $p^{(i)}, \epsilon^{(i)}$ that is bilinear in each of the $\epsilon^{(i)}$ and obeys the same gauge redundancy as the photons 
\eqref{gaugeph},  in addition to the constraints listed above.  

\subsection*{Gluons} 
The gluon S-matrix is characterised by the adjoint valued polarisations $\epsilon^{(i), a}_\mu$ and momenta $p^{(i)}_\mu$ where $a$ indicates the adjoint colour index. The general analysis in \cite{Chowdhury:2020ddc} was, in principle, applicable to any lie group $G $ but for concreteness, we will restrict to $SO(N)$ in this note. The gluon S-matrix, referred to as $\mathcal{S}^{\tg} (p^{(i)}, \epsilon^{(i),a})$, must be invariant under linearised gauge group transformations and non-constant infinitesimal gauge transformations $\zeta^{(i),a}$ 
\be
\epsilon_{\mu}^{(i),a} \to R^a_{\, b}\, \epsilon_{\mu}^{(i),b}, \qquad \epsilon_{\mu}^{(i),a} \to \epsilon_{\mu}^{(i),a} +p_\mu^{(i)} \zeta^{(i),a}.
\ee

To linear order, the authors of \cite{Chowdhury:2020ddc} interpreted the adjoint valued polarisation as $\epsilon_{\mu}^{a}=\epsilon_{\mu} \otimes \tau^a$ and enumerated the gluon S-matrix as the sum of products,
\be\label{tensor}
\mathcal{S}^{\tg}(\epsilon_{\mu}^{(i),a},p_{\mu}^{(i)}) =  \, \mathcal{S}^{\tv} (\epsilon_{\mu}^{(i)},p_{\mu}^{(i)})\, \mathcal{S}^{\ts, SO(N)}(\tau^{(i),a},0)+\ldots 
\ee 
i.e we can interpret the gluon S-matrix as the ``product" of photon S-matrix and S-matrix of adjoint scalar particles at zero momentum. The resulting S-matrix is invariant under the following transformations,
\be
\tau^{(i),a}\to R^a_{\, b} \, \tau^{(i),b},\qquad\qquad  \epsilon_{\mu}^{(i)} \to \epsilon_{\mu}^{(i)} +p_\mu^{(i)} \zeta^{(i)}.
\ee

Finally since we are considering scattering of identical particles, the S-matrix must also be invariant under the permutation group $S_4$ \cite{Chowdhury:2019kaq}.

The structure of this note is as follows. In section \ref{smam}, we review the module structure of the S-matrix as a consequence of  imposing $S_4$ invariance. In section \ref{eup}, we explain the module counting as being encoded in a multi particle partition function computing the symmetric product of four single letter partition functions. We explain how to project out the parity violating sector in the multi particle partition function and record the formulae we use to do so. In section \ref{epopied}, we explicitly compute the parity violating partition function for scalars, gravitons and gluons. Our parity violating counting is in agreement with the observations in \cite{Chowdhury:2019kaq, Chowdhury:2020ddc}.   
 
\section{S matrices as a module}\label{smam}

Invariance under the permutation symmetry group $S_4$ is done in two steps because the mandelstam invariants $s, t$ are invariant under the normal subgroup $\Z_2 \times \Z_2$ of $S_4$.\footnote{Refer Appendix \ref{s3review} for the nomenclature to be used throught the paper} 
\begin{equation} \label{permgp}
\frac{S_4}{\left( \Z_2 \times \Z_2 \right)}=S_3.
\end{equation}
First we impose invariance under $\Z_2 \times \Z_2 $- 
this is the abelian sub group of $S_4$ which is generated by the simultaneous exchange of two pairs of elements while $S_3$ is the permutation group of three objects. This normal subgroup is labelled by the abelian charges under the action of its generators $(P_{12}P_{34}, P_{13}P_{24})$, where $P_{ij}$ denotes the exchange $i \leftrightarrow j$. It has four irreducible one dimensional representations which are labelled by their abelian charges $(P_{12}P_{34}, P_{13}P_{24}, P_{14}P_{23})$.

\subsubsection*{Quasi-invariant S-matrix}
The state with the charge $(+++)$ is termed ``quasi-invariant" S-matrix \cite{Chowdhury:2019kaq, Chowdhury:2020ddc}. In practice it is obtained by applying the following projector on the gauge invariant, Lorentz-invariant tensor structure of polarisation tensors and momenta.

\be\label{pppproj} 
\frac{1+P_{12}P_{34}}{2} \frac{1+P_{13}P_{24}}{2} \frac{1+P_{14}P_{23}}{2}.\ee

Upon imposing $\Z_2 \times \Z_2$ invariance, the space of structures structures obtained for scalar ($\CM^{\rm scalar}$), photons ($\CM^{\rm photon}$)and gravitons ($\CM^{\rm graviton}$) have charges $(+++)$ only. In general, we are interested in classification of such quasi-invariant structures (and eventually S-matrices) that are obtained from local lagrangians of arbitrarily high derivative orders. In \cite{Henning:2017fpj, Chowdhury:2020ddc}, the authors were able to map this task to a linear algebra problem. Since the mandelstam invariants are also quasi-invariant, we see that the space of analytic quasi-invariant structures form a finite-dimensional module over the ring of functions $(s,t)$ \cite{Henning:2017fpj, Chowdhury:2020ddc}. Polynomials of $(s,t)$ cannot form a field since it has no multiplicative inverse. The scalar, photon and graviton module has finite number of module generators ($g_i$) and all the elements of the module can be written as a linear combinations of the generators and the ring elements $r_i$ as $\sum_i g_i \cdot r_i$. The module generators can be thought of as being obtained from the local gauge-invariant and Lorentz invariant lagrangians of the lowest derivative order. The other elements in the module correspond to higher derivative lagrangians. In dimensions greater than four, the modules are generated freely- every element of the module is a unique combination of the generators and ring element. In the language of local lagrangians, this means that the local lagrangians corresponding to the elements of the module are obtained by sprinkling contracted derivatives on the lowest derivative local lagrangian- these are often referred to as ``descendants". However in four dimensions, the authors of \cite{Chowdhury:2019kaq} observed that the photon and graviton modules are not generated freely- there are relations $\sum_i g_i \cdot r_i =0$ and classified them. In the language of local lagrangians, S-matrices generated by the descendants of generator lagrangians, are not linearly independent.   
         
\subsubsection*{Non-quasi-invariant S-matrix}\label{nqismat}

We can also consider states which have the charges $\{+--, -+-, --+\}$ under $\Z_2 \times \Z_2$. They can be generated by the projector 

\be\label{pppproj2} 
\frac{1\pm P_{12}P_{34}}{2} \frac{1 \pm P_{13}P_{24}}{2} \frac{1\pm P_{14}P_{23}}{2}.\ee
with not all + signs simultaneously. Because of the coset structure, $S_3$ acts non-trivially on elements of $\Z_2 \times \Z_2$ and in particular relates the non-invariant states to one another. In \cite{Chowdhury:2020ddc}, it was shown that a non-invariant state can transform either in $\bf{3}$ or as $\bf{3_A}$. We don't encounter such states when characterizing the photon, scalar and the graviton module. The context in which this becomes relevant is when we consider the gluon S-matrix as being generated by the product of colour module and the photon module \eqref{tensor}. The scalar module is generated by $\CM^{\rm colour}$. Since the derivatives do not act on the scalar module, we can think of them forming a vector space $\CV^{\rm scalar}$ over ${\mathbb C}$. These generators are nothing but colour structures at zero order in derivatives and hence generate the colour module freely. The total gluon module is a sum of two contributions,
\bea
\CM^{\rm gluon}&=& (\CM^{\rm photon} \otimes \CV^{\rm scalar}) \oplus \CM^{\rm non-inv},\nonumber\\
\eea
where $(\CM^{\rm photon} \otimes \CV^{\rm scalar})$ generates $\CM^{\rm inv}$, the tensor product of modules which are separately quasi-invariant. $\CM^{\rm non-inv}$, on the other hand, is a product of non-quasi-invariant colour and photon modules in a way such that the product is quasi-invariant. This was acheieved in \cite{Chowdhury:2020ddc} as follows, consider a possible non-quasi-invariant photon module  having states $\{|p^{(1)}\rangle, |p^{(2)}\rangle, |p^{(3)}\rangle\}$ which carry the charges $\{+--, -+-, --+\}$ respectively. We now tensor it with the corresponding colour non-invariant module carrying the same charges  $\{|s^{(1)}\rangle, |s^{(2)}\rangle, |s^{(3)}\rangle\}$ such that the tensor product has the charge $(+++)$. Thus in equations, the orbit of the quasi-invariant gluon module constructed in this manner is given by, 
\be
|p^{(1)}\rangle|s^{(1)}\rangle, |p^{(2)}\rangle|s^{(2)}\rangle, |p^{(3)}\rangle|s^{(3)}\rangle.
\ee  
In the rest of the paper, we will denote this operation by $\tilde{\otimes}$. The $S_3$ transformation property of this gluon module is obtained from the parent modules. If both of them transform in $\bf{3}$ or $\bf{3_A}$, the quasi invariant generator transforms in $\bf{3}$, while if just one of the parent module transforms in $\bf{3_A}$, the gluon module also transforms in $\bf{3_A}$\footnote{The reducible and irreducible representations of $S_3$ has been reviewed in appendix \ref{s3review}}. 

S-matrices are finally obtained by  projecting the elements of the module obtained in this way onto $S_3$ singlets. An efficient way of characterizing the space of S-matrices is to characterise the generators of the module $g_i$ by its $S_3$ irreps. The process of projecting the module element onto the $S_3$ singlet is then achieved in the following manner
\begin{equation} \label{cgi} 
\CS^\alpha(p_i,\epsilon_i)=\sum_{\sigma\in S_3} M^{\alpha, \sigma}(p_i,\epsilon_i)=\sum_{J\in L} \sum_{\sigma\in S_3} P_{{\bf R}}^{J, \sigma}(s,t)e_{{\bf R}}^{J, \alpha, \sigma}(p_i,\epsilon_i).
\end{equation} 
Where $\alpha \in \{\ts,\tg,\tt,\tv\}$ denotes the species of particle, $e_{\bf R}^{J, \alpha, \sigma}(p_i,\epsilon_i)$ is a module generator for $\alpha$ with a particular $S_3$ irrep ${\bf R}$, $P_{{\bf R}}^{J, \sigma}(s,t)$ is a ring element transforming in the same irrep ${\bf R}$, the superscript $\sigma$ denotes the action of $S_3$ permutation and $L$ represents the basis of generators.
\subsection{Partition function for the local module}
Since the module generators can be organised by their derivative order and $S_3$ properties, a very convenient way of characterising the number of linearly independent S-matrices that are obtained in this manner at each derivative order is through a partition function. Equation \eqref{cgi} tells us how the partition function should be enumerated. In order to project onto S-matrices, i.e $S_3$ singlets, we ``dot product" each module generator, transforming in a irreducible representation $\bf{R}$ of $S_3$,  with a polynomial of $(s,t)$, transforming in the same representation. If the derivative order of the generator $e^{J, \alpha}_{\bf R}$ is $n$, we get the following contribution to the partition function from the S-matrix  \eqref{cgi}
\be
Z_{e_{\bf R}}(x)=Z_{\bf R}(x) x^{n}, 
\ee
where $Z_{\bf R}(x)$ is the partition function for polynomials of $(s,t)$ transforming in representation ${\bf R}$ (see equation \eqref{partfnapp} for explicit expressions for partition functions for irreps of $S_3$). Hence following equations \eqref{cgi} and \eqref{partfnapp}, the partition function for the total S-matrix can be written down as, 
\be\label{pf-from-gen}
Z_{\rm S-matrix}=\sum_{J} Z_{\bf R}x^{n_{e^J_{\bf R}}}.
\ee
where we have summed over the contribution of all generators transforming in the three possible irreps of $S_3$ and the partition functions are defined in \eqref{Z-S3}. The partition functions for Photons, gravitons and gluons have been systematically enumerated in this manner in \cite{Chowdhury:2019kaq, Chowdhury:2020ddc}.

\section{Enumeration using Plethystics}\label{eup}
The local module is in one-to-one correspondence with equivalence class of local lagrangians, i.e local lagrangians upto field re-definitions and total derivatives \cite{Chowdhury:2019kaq, Chowdhury:2020ddc}. These lagrangians will be quartic polynomials of the field strength $F_{\mu\nu}$ for photons, Riemann tensor $R_{abcd}$ for gravitons (mostly) and adjoint field strength $\rm{Tr}(F_{\mu \nu})$ for gluons. Thus an alternative method of counting the module partition function can be formulated. We write down the single letter partition function which encodes number of operators that involve a single field and derivatives modulo equations of motion. 
\begin{eqnarray}\label{single}
i(x,y,z)&=&{\rm Tr}\,\,x^{\partial} y_i^{L_i}z_\alpha^{H_\alpha},
\end{eqnarray}
where $\partial$ is the number of derivatives and $x$ keeps track of it, $H_\alpha$ denotes the cartan of internal symmetry group $G$ while $L_i$ denotes the cartan of the charges of the angular momentum generators. In general it is a sum over characters of irreducible representations of $G$ and angular momentum graded by derivatives, which is encoded in a suitable polynomial of $x$.\footnote{Similar techniques were used in \cite{Sundborg:1999ue, Aharony:2003sx} to compute partition functions for gauge theories at weak coupling.} The character corresponding to a representation $l$ of a group $G$ is defined as 
\be
\chi_l(y)=\text{Tr}_l(g)=\sum_\alpha \langle \alpha | \prod_i y_i^{f_i} |\alpha\rangle.
\ee
where $g$ denotes an element of the group $G$ in representation $l$, $f_i$ denotes the cartan charges of the group $G$, $|\alpha\rangle$ denotes the basis vectors of the representation $l$ which are labelled by the weights $f_i$ corresponding to the irrep $l$. In \cite{Chowdhury:2019kaq, Chowdhury:2020ddc}, the authors chose the angular momentum generators to be $SO(D)$ and the resulting analysis did not differentiate between parity invariant and the parity violating contributions. The problem of enumerating linearly independent lagrangians at a given derivative order is similar to enumerating number of Lorentz and $G$ singlets that we can build out of four such identical operators, which are generated by the partition function \eqref{single}. For four identical particles, the relevant multi-particle function turns out to be, 

\bea\label{4-particle}
i^{(4)}(x,y,z)&=&\frac{1}{24}\Big(i^4(x,y,z) +6 i^2(x,y,z) i(x^2,y^2,z^2)+3i^2(x^2,y^2,z^2)\nonumber\\
&+&8i(x,y,z)i(x^3,y^3,z^3)+6i(x^4,y^4,z^4)\Big).\nonumber\\
\eea
This is sometimes referred to as plethystic exponential \cite{Gray:2008yu, Henning:2017fpj}.  

\bea\label{4-particle pleth}
i^{(4)}(x,y,z)&=&e^{\sum_n \frac{t^n i(x^n,y^n,z^n)}{n}}|_{t^4},\nonumber\\
\eea

where for four particle partition function, we look at the coefficient of $t^4$. We now integrate this over the Lorentz Haar measure and the $G$ Haar measure to project onto the singlets 

\be\label{singlet-proj}
I^D(x):=\oint  d\mu_{G}~ \oint  d\mu_{a}~  i^{(4)}(x,y,z)/\denom(x,y).
\ee  
Here $d\mu_{G}$ is the Haar measure associated with the group $G$ and $d\mu_{a}$ denotes the Haar measure corresponding to space time symmetry, and $\denom(x,y)$, projects out total derivatives from the set. When the space time symmetry is $SO(D)$ (where $D$ is even),
\be
\denom(x,y)=\frac{1}{\Pi_{i=1}^{D/2}\left(1-x y_i\right)\left(1-\frac{x}{y_i} \right)}.
\ee 

In \cite{Chowdhury:2019kaq, Chowdhury:2020ddc}  space-time symmetry was taken to be $SO(D)$ which correspond to computing the sum of parity invariant and parity violating sectors. The partition function obtained in this manner should be the same as $Z_{\rm S-matrix}$ given by \eqref{pf-from-gen}.

\subsection{Parity}
In order to project onto the parity invariant and parity violating sectors, we need to understand the group structure of $O(D)$, the group of matrices with $R^TR=1$ and $\text{Det}R= \pm1$. Recall that, parity acts as a reflection in $\R^D$ and hence the group $O(D)$ is a (semi) direct product of $SO(D)$ and this $\Z_2$ action in (even) odd dimensions. If we denote the parity element as $\mathcal{P}$, an element $g$ of $O(D)$ can be represented by cosets of the subgroup $SO(D)$,

\be 
g\in \{g_+ \cup g_-\}
\ee 
where $g_+ \in SO(D)$ and $g_- \in SO(D) \mathcal{P}$. This implies that for constructing the module that is parity invariant, we must ensure invariance under $O(D)$ rather than $SO(D)$. Consequently the cartan charges for the angular generators to be that of $O(D)$ rather than $SO(D)$ \cite{Henning:2017fpj}. This can be implemented in the following way. 
  
\be\label{singlet-proj-inv}
\begin{split}
	I^{D,~ inv}(x)&:= \frac{1}{2} \left(I^{D,~ even}(x) + I^{D,~odd}(x)\right),~I^{D,~non- inv}(x):= \frac{1}{2} \left(I^{D,~ even}(x) - I^{D,~odd}(x)\right),\qquad\\
	I^{D,~ even}(x)&= \oint  d\mu_{G}~ \oint  d\mu_{ +}~  \frac{i^{(4), +}(x,y,z)}{\denom_+(x,y)},~ 
	I^{D,~ odd}(x) =\oint  d\mu_{G}~ \oint  d\mu_{-}~  \frac{i^{(4), -}(x,y,z)}{\denom_-(x,y)}.
\end{split}
\ee  

where we have broken \eqref{singlet-proj} into two parts $I^{D,~ even}(x)$ and $I^{D,~ odd}(x)$, the parity even and parity odd pieces of $O(D)$. The quantities $\denom_+(x,y)$ and $\denom_-(x,y)$ respectively remove total derivatives when the cartans of space-time symmetry are chosen to be that of $SO(D)$ and $SO(D)\mathcal{P}$ respectively. For $I^{D,~ even}(x)$, the cartan charges in the single letter partition function \eqref{single} and Haar measure for Lorentz singlets is $SO(D)$- we denote this contribution by tracing over group elements $g_+$ in irreducible representations of $SO(D)$.  This was precisely evaluated in \cite{Chowdhury:2019kaq, Chowdhury:2020ddc}. The evaluation of $I^{D,~ odd}(x)$ is much more subtle and is different for odd and even dimensions. We must also normalise the corresponding Haar measures such that $\int d \mu_{\pm} =1$ where $d\mu_+$ and $d\mu_-$ respectively correspond to $SO(D)$ and $SO(D)\mathcal{P}$ Haar measures respectively. To set the conventions in the rest of the paper, when we talk about characters of $SO(D)$ and $SO(D)\mathcal{P}$, we will use $\chi_l^+(y)$ and $\chi_l^-(y)$ respectively.
\subsubsection{Odd $D$: O(2N+1)}\label{po2}

In odd dimensions ($D=2N+1$), parity is a linear map on spatial coordinates which takes vector to its image under reflection.
\be\label{defnparity}
\mathcal{P}: x^\mu \rightarrow -x^\mu,
\ee   

where $\mathcal{P} \in O(2N+1)$ and $\text{Det} \mathcal{P}=-1$, this is the vector representation of parity operator. For an arbitrary tensor, this implies that action of parity depends on the rank of the tensor. If the tensor has odd rank, it flips sign under parity while it does not if its rank is even. The representation matrix of the parity operator $\mathcal{P}$ is $\rho_{l}(\mathcal{P})= (-1)^{|\sum l_i|}\mathcal{I}$, for an arbitrary tensor representation of $SO(2N+1)$ denoted by $(l_1, l_2,\cdots l_{N})$, where $\mathcal{I}$ is identity matrix. As an example, we present how parity affects the single and multi-letter partition function for scalars and then present the general result that was derived in \cite{Henning:2017fpj}. Recall that for scalars the single letter partition function is a generating function for operators of the form $\partial_{\mu_1}\partial_{\mu_2}\cdots \phi$ modulo equations of motion $\partial^2 \phi=0$. In equations, the single letter partition function corresponding to $SO(2N+1)\mathcal{P}$ thus takes the form, 

\be
\begin{split}
	i^-_{\ts}(x,y)&={\rm Tr}\,\,x^{\partial} y_i^{L_i} \mathcal{P},\\
	&= x(1-x\,\chi^+_{\syng{1}}+x^2\, \chi^+_{\syng{2}}-x^3\, \chi^+_{\syng{3}}+x^4\, \chi^+_{\syng{4}}-x^5\, \chi^+_{\syng{5}}+\ldots), \nonumber \\
	&=\frac{x (1-x^2)}{\Pi_{i=1}^{N}\left(1+x y_i\right)\left(1+\frac{x}{y_i} \right)(1+x)}= x(1-x^2)\denom^O_-(x,y),
\end{split}
\ee   
where the $L_i$ are the cartan charges of $SO(2N+1)$, $ \chi^+_{\syng{2}\cdots}$ denotes the character corresponding to symmetric traceless tensor for $SO(2N+1)$ and the superscript $O$ in $\denom^O_-(x,y)$ denotes odd dimensions. Note that in comparison with \cite{Chowdhury:2019kaq}, here we have retained the scaling of individual $\phi$ field. We have also used that 
\be
\frac{1}{\Pi_{i=1}^{N}\left(1+x y_i\right)\left(1+\frac{x}{y_i} \right)(1+x)}=\sum (-1)^n x^n \chi^+_{\text{Sym}^n \syng{1}} (y),
\ee 
and the fact that we have to subtract out the traces from the symmetric product. Noting that $i^+_{\ts}(x,y)= x(1-x^2)\denom^O_+(x,y)$ where, $\denom^O_+(x,y)=\frac{1}{\Pi_{i=1}^{N}\left(1-x y_i\right)\left(1-\frac{x}{y_i} \right)(1-x)}$, we conclude that $\frac{i^-_{\ts}(x,y)}{x}= \frac{i^+_{\ts}(-x,y)}{-x}$. The $SO(2N+1)$ measure is unchanged under parity and we obtain, 
\be
\begin{split}\label{scalarodd}
	I_\ts^{D,~ odd}(x) &=\oint  d\mu_{G}~ \oint  d\mu_{-}~  i_\ts^{(4),-}(x,y,z)/\denom^O_-(x,y)=I_\ts^{D,~ even}(-x).
\end{split}
\ee 
The overall negative sign due to the scaling of $\phi$ in the single letter partition function does not matter since, the resulting four-particle partition function will be quartic in $\phi$. For instance, we could have chosen to define our single letter partition function as $\tilde{i}^+_{\ts}(x,y)= \frac{i^+_{\ts}(x,y)}{x}$ and hence $\tilde{i}^-_{\ts}(x,y)= \frac{i^-_{\ts}(x,y)}{x}$. This would not have affected the relation \eqref{scalarodd} except a division by an overall factor $x^4$. For a generic particle transforming in the irrep $(l_1,l_2,l_3,\cdots, l_N)$ of $SO(2N+1)$, the single letter partition function takes the form $$i^+_{gen}(x,y)=\sum_n x^{n+\sum_i l_i} \chi^+_{(n+l_1,l_2, l_3, \cdots)}(y),$$
which encodes the tower of operators constructed from symmetrized derivatives on the representation $(l_1,l_2,l_3,\cdots, l_N)$ modulo equations of motion and relevant identities (for e.g., bianchi identities for Riemann polynomials). It is clear that the argument presented above goes through and we obtain, 
\be\label{singlet-proj-odd}
\begin{split}
	I^{D,~ inv}(x)&:= \frac{1}{2} \left(I^{D,~ even}(x) + I^{D,~even}(-x)\right), \qquad 
	I^{D,~ non-inv}(x):= \frac{1}{2} \left(I^{D,~ even}(x) - I^{D,~even}(-x)\right).\\
\end{split}
\ee 
In \cite{Chowdhury:2019kaq}, this was the observation from the local module perspective as well since the explicit construction of parity violating  module involve Levi-Civita tensors- the parity violating partition function contribution to the plethysic are odd polynomials in $x$. In odd dimensions the Levi-Civita tensors have odd number of indices and explicit local module counting showed that overall derivative scaling is odd. Intuitively it can be understood as follows, the basic building blocks of photon and graviton modules are quartic polynomials of $F_{\mu \nu}$ and $R_{\mu\nu\alpha\beta}$ which overall have even number of derivatives and polarisations. Hence, a Lorentz singlet, at the lowest order in derivatives, involving quartic polynomials of field strength and Riemann tensors with Levi-Civita tensors must involve odd number of derivatives to saturate the indices of the Levi-Civita tensor.  

\subsubsection{Even $D$: $O(2N)$}\label{po1}
 
In even dimensions ($D=2N$), the action of parity as defined in \eqref{defnparity} is ambiguous- it coincides with a rotation. Consequently the action is chosen in a different manner, as a reflection orthogonal to a hyperplane in $\R^{2N}$ with the trade off that $\mathcal{P}$ no longer commutes with rotations in $SO(2N)$. The vector representation of $\mathcal{P}$ is given by 
$$\rho_{\syng{1}} (\mathcal{P})= \text{Diag}\{1,1,\cdots,1,-1\}.$$

The resulting analysis of characters and plethystic exponential has been presented in Appendix C of \cite{Henning:2017fpj} and we summarise them here. For a generic irreducible representation labelled by $l \equiv (l_1, l_2,\cdots l_{N-1},0)$, the parity odd characters are \cite{Henning:2017fpj, 10.2307/j.ctv3hh48t.1}

\be\label{poch} 
\chi_l^+ (y)= \chi^{SO(2N)}_l (y), \qquad \chi_{(l_1, l_2,\cdots l_{N-1},0)}^- (y)= \chi^{Sp(2N-2)}_{(l_1, l_2,\cdots l_{N-1})} (\tilde{y}),
\ee  
where $\tilde{y}= (y_1,y_2,\cdots, y_{N-1})$ and $Sp(2N-2)$ denotes the Symplectic group labelled by the same subset of cartan charges. As explained in \cite{Henning:2017fpj}, the evaluation of the character $\chi^-_l$, requires us to identify the states which remain invariant under parity given a particular highest weight state. The action of parity in even dimensions turns out to be ``folding" the root system of $SO(D)$ to get parity symmetrized new set of simple roots which precisely coincide with the root system of $Sp(D-2)$. The second subtlety arises in the evaluation of the trace $\text{Tr}_l((g_+ \mathcal{P})^n)$ which is needed for the multi-particle partition function. 
\be\label{plethodd}
\begin{split}
	\text{Tr}_l((g_+ \mathcal{P})^n)= \begin{cases}
		\chi_l^{SO(2N)}(\bar{y}^n), \qquad n=2k\\
		\chi_l^{Sp(2N-2)}(\tilde{y}^n), \qquad n=2k+1 
	\end{cases}
\end{split}
\ee  
where $\bar{y}= (y_1,y_2,\cdots, y_{N-1},1)$. Let us illustrate these features explicitly with an example. The defining representation of the parity odd element is given by $\rho_{\syng{1}} (g_+ \mathcal{P})$, where $g_+ \in SO(D)$

\begin{equation}\label{examplevec}
\begin{split}
\text{Tr}_{\syng{1}}((g_+ \mathcal{P}))\sim&\left( \begin{matrix}
c_{\theta_1} & s_{\theta_1} &0 &\cdots &0  \\
-s_{\theta_1} & c_{\theta_1} &0 &\cdots &0\\
0 & 0 & c_{\theta_2} &\cdots &0 \\
0 & 0 &-s_{\theta_2} &c_{\theta_2}\cdots &0\\
\cdots&\cdots& \cdots& \cdots&\cdots\\
0 & 0 &\cdots&c_{\theta_n} &s_{\theta_n}\\
0 & 0 &0 &\cdots &c_{\theta_n}\\
\end{matrix}\right) \left( \begin{matrix}
1 & 0 &0 &\cdots &0  \\
0 & 1 &0 &\cdots &0\\
0 & 0 & 1 &\cdots &0 \\
0 & 0 &0 &1\cdots &0\\
\cdots&\cdots& \cdots& \cdots&\cdots\\
0 & 0 &\cdots&1 &0\\
0 & 0 &0 &\cdots &-1\\
\end{matrix}\right),\\
&\sim\left( \begin{matrix}
e^{i \theta_1} & 0 &0 &\cdots &0  \\
0 & e^{-i \theta_1} &0 &\cdots &0\\
0 & 0 & e^{i \theta_2} &\cdots &0 \\
0 & 0 &0 &e^{-i \theta_2}\cdots &0\\
\cdots&\cdots& \cdots& \cdots&\cdots\\
0 & 0 &\cdots&1 &0\\
0 & 0 &0 &\cdots &-1\\
\end{matrix}\right)
\end{split}
\end{equation}

where in the second line we have made a diagonalisation. The character for the defining representation is 
\be 
\chi^-_{\syng{1}}(y)= \text{Tr}_{\syng{1}}(g_+ \mathcal{P}) = \sum^{N-1}_{i=1} (y_i+1/y_i).  
\ee  
We note that the character depends on $N-1$ torus coordinates $\tilde{y}= (y_1,y_2, \cdots y_{N-1})$ as advertised. From \eqref{examplevec}, we can also see that  
\begin{equation}
\text{Tr}_{\syng{1}}((g_+ \mathcal{P})^n)=\begin{cases}
\sum^{N-1}_{i=1} (y^n_{i}+1/y^n_i)+2, \qquad n=2k\\
\sum^{N-1}_{i=1} (y^n_{i}+1/y^n_i), \qquad n=2k+1 
\end{cases}
\end{equation}
in accordance with \eqref{plethodd}. The plethystic exponential of the parity odd character $\chi_l^- (y)$ therefore becomes \cite{Henning:2017fpj}
\begin{equation}
\begin{split}
e^{\sum_n \frac{t^n \text{Tr}_l((g_+ \mathcal{P})^n)}{n}} = e^{\sum_n \frac{t^n \chi_l^{Sp(2N-2)}(\tilde{y}^n) + \frac{t^{2n}}{2}(\chi_l^{SO(2N)}(\bar{y}^{2n})-\chi_l^{Sp(2N-2)}(\tilde{y}^{2n}))}{n}}.
\end{split}
\end{equation}  
In summary, we can now write down the relevant formulae for the parity odd and parity even plethystic in even dimensions.The parity even contribution is given by 
\begin{equation}\label{parity1}
\begin{split}
I^{2N,~ even}(x) &=\oint  d\mu_{G}~ \oint  d\mu_{+}~  i^{(4), +}(x,y,z)/\denom^E_+(x,y).\\
i^{(4), +}(x,y,z) &=\frac{1}{24}\Big(i^{SO(2N)}(x,y,z)^4 +6 i^{SO(2N)}(x,y,z)^2 i^{SO(2N)}(x^2,y^2,z^2)+3i^{SO(2N)}(x^2,y^2,z^2)^2\\
&+8i^{SO(2N)}(x,y,z)i^{SO(2N)}(x^3,y^3,z^3)+6i^{SO(2N)}(x^4,y^4,z^4)\Big),\\
\denom^E_+(x,y)&=\frac{1}{\Pi_{i=1}^{N}\left(1-x y_i\right)\left(1-\frac{x}{y_i} \right)}.
\end{split}
\end{equation}
where the measure $d\mu_{+}$ is the Haar measure corresponding to $SO(2N)$. While the parity odd contribution is given by,

\begin{equation}\label{parity2} 
\begin{split}
I^{2N,~ odd}(x) &=\oint  d\mu_{G}~ \oint  d\mu_{-}~  i^{(4), -}(x,y,z)/\denom^E_-(x,y).\\
i^{(4), -}(x,y,z) &=\frac{1}{24}\Big(i^{Sp(2N-2)}(x,\tilde{y},z)^4 +6 i^{Sp(2N-2)}(x,\tilde{y},z)^2 i^{SO(2N)}(x^2,\bar{y}^2,z^2)+3i^{SO(2N)}(x^2,\bar{y}^2,z^2)^2\\
&+8i^{Sp(2N-2)}(x,\tilde{y},z)i^{Sp(2N-2)}(x^3,\tilde{y}^3,z^3)+6i^{SO(2N)}(x^4,\bar{y}^4,z^4)\Big),\\
\denom^E_-(x,y)&= e^{\sum_n \frac{t^n \chi_{\syng{1}}^{Sp(2N-2)}(\tilde{y}^n) + \frac{t^{2n}}{2}(\chi_{\syng{1}}^{SO(2N)}(\bar{y}^{2n})-\chi_{\syng{1}}^{Sp(2N-2)}(\tilde{y}^{2n}))}{n}}=\frac{(1-x^2)}{\Pi_{i=1}^{N-1}\left(1-x y_i\right)\left(1-\frac{x}{y_i} \right)}.
\end{split}
\end{equation}
where we have used, 
 \bea\label{4-particle pleth-noninv}
 i^{(4),-}(x,y,z)&=&e^{\sum_n \frac{t^n i^-(x^n,\tilde{y}^n,z^n)+\frac{t^{2n}}{2}\left(i^+(x^{2n},\bar{y}^{2n},z^{2n})-i^-(x^{2n},\tilde{y}^{2n},z^{2n})\right)}{n}}|_{t^4}.\nonumber\\
 \eea
The measure $d\mu_{-}$ is the Haar measure corresponding to $Sp(2N-2)$. Note that in these formulae, we denote the charges of the cartans of $SO(D)$ and $Sp(D-2)$ by the same variable $y_i$ since parity acts on $SO(D)$ by folding its root system to give $Sp(D-2)$. In the subsequent sections we evaluate \eqref{parity1} and \eqref{parity2} for photons, gravitons and gluons. 

\section{Enumerating parity odd plethystic in even dimensions}\label{epopied}
In this section we work out the explicit parity odd single letter partition function and the associated counting for parity violating four particle module following subsection \eqref{po1}. We  first summarise the different identities and expressions for the characters of the different representations needed for evaluating the Haar integral over partition functions. The irreducible representations of the scalar, symmetric two tensor and the anti symmetric two tensor representations are respectively represented via young tableaux in the following manner

\begin{eqnarray}
\cdot,\, {\tyng(2)},\, {\tyng(1,1)}
\end{eqnarray}
The Symplectic group differs from the special orthogonal group in the fact in the presence of the antisymmetric tensor invariant $f_{ab}$ instead of the symmetric invariant tensor $\delta_{ab}$ \cite{cvitanovic}, which basically implies that the trace appears in the antisymmetric sector rather than the symmetric sector, which is the case for $SO(D)$. To be precise, consider the following tensor products of the vector representation 
\begin{eqnarray}
S^2 {\tyng(1)}, \Lambda^2 {\tyng(1)}
\end{eqnarray}  
We note the following identity for tensor product of vector representation of $Sp(D)$ groups, 

\begin{equation}
\begin{split}
S^2 {\tyng(1)}= {\tyng(2)},~~ \Lambda^2 {\tyng(1)}= {\tyng(1,1)} \oplus \cdot 
\end{split}
\end{equation}
We note the difference compared to the $SO(D)$ case where it was the symmetric sector which decomposed into the traceless and trace part. \footnote{In general all these observations can be summed up in the ``negative dimensionality theorem" stated in \cite{cvitanovic} (where we refer to the more mathematically inclined reader for details)- which states that the group theoretic relation $SO(n)= \bar{Sp(-n)}$}. For later use we note the respective symmetric traceless and traceless adjoint characters for the two groups \cite{Balantekin:2000vn, Balantekin:2001id}.

\begin{eqnarray}\label{spsocharacters}
\chi^{Sp(D)}_{(n,0,0,0\cdots,0)} (y)= h_n(y_i, y^{-1}_i), \qquad \chi^{SO(D)}_{(n,0,0,0\cdots,0)} (y)= h_n(y_i, y^{-1}_i)- h_{n-2}(y_i, y^{-1}_i), \nonumber\\
\chi^{Sp(D)}_{(1,1,0,0\cdots,0)} (y)= a_2(y_i, y^{-1}_i)-a_0(y_i, y^{-1}_i), \qquad \chi^{SO(D)}_{(1,1,0,0\cdots,0)} (y)= a_2(y_i, y^{-1}_i),\nonumber\\
\end{eqnarray}
where the functions $h$ and $a$ are defined as follows, 

\begin{eqnarray}
\frac{1}{\Pi_{i=1}^{\frac{D}{2}}(1-y_i x)(1- x/y_i)}= \sum_n h_n(y_i, y^{-1}_i) x^n, \qquad \Pi_{i=1}^{\frac{D}{2}}(1-y_i x)(1- x/y_i)= \sum_n (-1)^n a_n(y_i, y^{-1}_i) x^n. \nonumber\\
\end{eqnarray}

\subsection{Scalars}

 We begin by recalling the single letter partition function for scalars and how it is obtained. The technique for obtaining this will be crucial for obtaining the objects that appear in the parity odd module evaluation. Following \cite{Chowdhury:2019kaq}, we can write the single letter partition function as follows. 

\begin{eqnarray} \label{sodscalar}
i^{SO(D)}_{\ts}(x,y)&=& x(1+x\,\chi^+_{\syng{1}}+x^2\, \chi^+_{\syng{2}}+x^3\, \chi^+_{\syng{3}}+x^4\, \chi^+_{\syng{4}}+x^5\, \chi_{\syng{5}}+\ldots \nonumber), \nonumber\\
&=&x (1-x^2)\denom^E_+(x,y),
\end{eqnarray}
where $ \chi^+_{\syng{2}\cdots}= \chi^{SO(D)}_{\syng{2}\cdots}$ denotes the symmetric traceless tensor for $SO(D)$ and we have used \eqref{spsocharacters}. The only parity violating scalar S-matrix module occurs in $D=3$ and is given by \cite{Chowdhury:2019kaq}
\begin{eqnarray}
\epsilon^{\mu\nu\alpha}k^1_{\mu}k^2_{\nu} k^3_{\alpha},	
\end{eqnarray}
which transforms in ${\bf{1_A}}$, in addition to the usual parity invariant scalar module which transforms in $\bf{1_S}$. This parity violating local module  and the usual parity invariant local module has the contributions to the partition function
\be Z^{D=3, non-inv}_\ts=\frac{x^{13}}{(1-x^4)(1-x^6)},~~Z^{ D=3, inv}_\ts=\frac{x^4}{(1-x^4)(1-x^6)}.\ee
For the direct plethystic evaluation, we get from \eqref{parity1}, \cite{Chowdhury:2019kaq} 
\be
I_\ts^{D=3,~ even}(x) = \frac{x^4(1+x^9)}{(1-x^4)(1-x^6)}.
\ee   
We see perfect agreement with \eqref{singlet-proj-odd}. Note that in \cite{Chowdhury:2019kaq}, the authors evaluated basically \eqref{parity1} and it was also conjectured that there are no parity violating module in higher dimensions as well. Intuitively this is obvious from the structure of the local module since in higher dimensions, there are no vectors to contract to make a Lorentz singlet. This implies that for even dimensions, \eqref{parity2} must be the same as \eqref{parity1}. To compute \eqref{parity2}, we need to sum the following series  
\begin{eqnarray} \label{spdscalar}
i^{Sp(2N-2)}_{\ts}(x,y)&=& x(1+x\,\chi^-_{\syng{1}}+x^2\, \chi^-_{\syng{2}}+x^3\, \chi^-_{\syng{3}}+x^4\, \chi^-_{\syng{4}}+x^5\, \chi^-_{\syng{5}}+\ldots), \nonumber \\
&=&\frac{x}{\Pi_{i=1}^{D/2-1}\left(1-x y_i\right)\left(1-\frac{x}{y_i} \right)}=\frac{x}{(1-x^2)}\denom^E_-(x,y),\\
\end{eqnarray}
where $ \chi^-_{\syng{2}\cdots}=\chi^{Sp(D-2)}_{\syng{2}\cdots}(\tilde{y})$ denotes the symmetric tensor for $Sp(D-2)$ and we have used \eqref{spsocharacters}. We can now evaluate the multi-particle partition function \eqref{parity2}. For the case $D=4$ we can explicitly do the numerical integral following the techniques in \cite{Chowdhury:2019kaq} and we obtain 
\begin{eqnarray}
I^{D=4, even}_\ts (x) = I^{D=4, odd}_\ts (x)= \frac{x^4}{(1-x^4)(1-x^6)}.
\end{eqnarray}  
Beyond a certain number of dimensions, the numerical integrals are difficult to perform and we perform a large $D$ analysis to show that the parity odd contribution remains zero in higher dimensions as well. We refer to appendix \ref{larged} for the technical details and in the main text we report the expected results, 

\begin{eqnarray}\label{largedscalar}
I^{D, even}_\ts (x) = I^{D, odd}_\ts (x)= \frac{x^4}{(1-x^4)(1-x^6)}.
\end{eqnarray} 

In summary, we have shown that there exists no parity violating contact interactions contributing to 2 to 2 tree level scattering of scalars for $D>3$.

\subsection{Photons}\label{photons}
We now move on to the more non trivial case of evaluating the module for parity violating photon S-matrices. In \cite{Chowdhury:2019kaq}, the authors classified the module generators for analytic S-matrices which contributed to 4-particle scattering of photons. Let us briefly summarise the findings. 

There are three generators of the photon module for $D\geq 8$ dimensions. Below $D=8$, the counting varied from dimensions to dimensions because of presence of parity violating structures contributing to scattering, unique to specific dimensions. In $D \geq 8$, two of the three module generators transform in $\bf{3}$ and the other one in $\bf{1_S}$- all of them are parity invariant and the local lagrangians generating the modules are given by 
\begin{equation}\label{lagstruct}
{\rm Tr}(F^2){\rm Tr}(F^2), ~~~~~ {\rm Tr}(F^4), ~~~~~~ -F^{ab} \partial_a F^{\mu\nu} \partial_b F^{\nu\rho} F^{\rho \mu}.
\end{equation} 
The corresponding generators are given by, 
\bea\label{photon-lag}
E^{(1)}_{{\bf 3}, 1} &=& 8{\rm Tr}(F_1F_2){\rm Tr}(F_3F_4),~~~~
E^{(1)}_{{\bf 3}, 2} = 8{\rm Tr}(F_1F_3F_2F_4),\nonumber\\
E_{\bf S}&\simeq& -6F^{ab}_{1}  \partial_a F_{2}^{\mu\nu} \partial_b F_{3}^{\nu\rho} F_{4}^{\rho \mu}|_{\Z_2\times \Z_2}\nonumber\\
&=& 6\left(-F^{ab}_{1}  \partial_a F_{2}^{\mu\nu} \partial_b F_{3}^{\nu\rho} F_{4}^{\rho \mu} -F^{ab}_{2}  \partial_a F_{1}^{\mu\nu} \partial_b F_{4}^{\nu\rho} F_{3}^{\rho \mu}-F^{ab}_{3}  \partial_a F_{4}^{\mu\nu} \partial_b F_{1}^{\nu\rho} F_{2}^{\rho \mu}-F^{ab}_{4}  \partial_a F_{3}^{\mu\nu} \partial_b F_{2}^{\nu\rho} F_{1}^{\rho \mu}\right). \nonumber\\
\eea
where the first subscript denotes the representation of $S_3$ in which the generator transforms, the superscript indicates the orbit element. The other two orbit elements can be obtained from $E^{(1)}_{{\bf 3}, i}$ by cyclic permutations. From now on, for brevity, we will not explicitly write down the full $\Z_2\times\Z_2$ projection of the module generator.  The contribution to these modules to partition function is 
\be I^{D,~inv}_{\tv}(x)=2x^8 Z_{{\bf 3}}+x^{10} Z_{{\bf 1_s}},\ee
where the partition functions are given in eqn \eqref{partfnapp}. 
For the lower dimensions, in addition to these parity invariant generators (and for $D=4$, reduction in the number of parity invariant generators), there were parity violating generators appearing at each dimension. For the odd dimensions $D=5,7$, the contribution of the parity violating module to the total counting is easy to determine, the local parity violating generators in odd dimensions have odd number of derivatives. 

For $D=6$, evaluation of \eqref{parity1} and comparison with large $D$ module structure, led to the conclusion that there was a single parity violating local module and it transformed in $\bf{1_A}$ representation of $S_3$. The module is generated by the local lagrangian,
\be\label{parity-odd-6d}
F^{ab}*(\partial_a F\wedge \partial_b F \wedge F),
\ee
with the generator 
\be
O^{D=6}_{\bf A}=F_1^{ab}*(\partial_a F_2\wedge \partial_b F_3 \wedge F_4)|_{\Z_2 \times \Z_2},
\ee 
and it's contribution to the partition function is  $I^{D=6,~non-inv}_{\tv}(x)=x^{10}Z_{{\bf A}}$.

In $D=4$, things were more subtle because of presence of relations in the module, i.e they were not freely generated \cite{Chowdhury:2019kaq}. There are two parity violating generators transforming in ${\bf 3}$ and ${\bf 1_S}$ respectively.

\be \label{Lagrangiansdfpo}
*(F\wedge F){\rm Tr}(F^2),\qquad \qquad \varepsilon_{\mu\nu\rho\sigma}F^{\mu\nu} \partial^\rho F^{ab}\partial^\sigma F^{bc} F^{ca}.
\ee
The corresponding generators are 
\be\label{parity-odd-4d} 
\begin{split}
	O_{\bf 3}^{D=4,(1)}&\equiv 2*(F_1\wedge F_2){\rm Tr}(F_3 F_4)|_{\Z_2\times \Z_2},\qquad
	O_{\bf S}^{D=4}\equiv 6\varepsilon_{\mu\nu\rho\sigma}F_1^{\mu\nu} \partial^\rho F_2^{ab}\partial^\sigma F_3^{bc} F_4^{ca}|_{\Z_2\times \Z_2},
\end{split}
\ee   
where $O_{\bf 3}^{D=4,(2)}, O_{\bf 3}^{D=4,(3)}$ can be constructed from $O_{\bf 3}^{D=4,(1)}$ by cyclic permutations. These local modules have relations amongst them so that the final contribution of these modules to the partition function is  $I^{D=4,~non-inv}_{\tv}(x)=x^8Z_{{\bf 3}}+x^{10}Z_{{\bf S}}-(x^{12}+x^{14})Z_{{\bf S}}$\footnote{See eq (5.53) of \cite{Chowdhury:2019kaq}}. In the following subsections, we explicitly calculate the parity odd contribution to the plethystic and confirm these module structures and properties along with the relations.

\subsubsection{Parity odd plethystic evaluation}
To determine the parity odd contribution to the partition function, we need to evaluate \eqref{parity2}. For $D\geq 8$, we use the \eqref{spsocharacters} to evaluate the $Sp(D-2)$ characters required in order to compute \eqref{parity2}. We note that analogous to the case of scalars, the evaluation of \eqref{parity1} has been done in \cite{Chowdhury:2019kaq}. Recall that the single letter partition function relevant for the purpose of computing \eqref{parity1} is given by, 
\begin{eqnarray}\label{pfph1}
i^{SO(D)}_{\tv}(x,y)&=& x^2\,\chi_{\syng{1,1}}+x^3\, \chi_{\syng{2,1}}+x^4\, \chi_{\syng{3,1}}\ldots \nonumber \\
&=& i^{SO(D)}_{\ts}(x,y)(\chi_{\syng{1}}) - \left(\frac{i^{SO(D)}_{\ts}(x,y)}{x} -1\right) -x i^{SO(D)}_{\ts}(x,y), \\
\end{eqnarray}
where $i^{SO(D)}_{\ts}(x,y)$ is defined in \eqref{sodscalar}. For evaluating \eqref{parity2}, we need $i^{SO(D)}_{\tv}(x,\bar{y})$ where $\bar{y}=(y_1,y_2,\cdots,y_{\frac{D}{2}-1},1)$. From equation \eqref{pfph1}, 

\begin{eqnarray}\label{pfph2}
i^{SO(D)}_{\tv}(x,\bar{y})&=& \frac{i^{SO(D-2)}_{\ts}(x,\tilde{y})}{1-x^2}(\chi^{SO(D)}_{\syng{1}}(\bar{y})) - \left(\frac{i^{SO(D-2)}_{\ts}(x,\tilde{y})}{x(1-x^2)} -1\right) -\frac{x i^{SO(D-2)}_{\ts}(x,\tilde{y})}{(1-x^2)},\nonumber\\
\chi^{SO(D)}_{\syng{1}}(\bar{y})&=& \chi^{SO(D-2)}_{\syng{1}}(\tilde{y})+2,
\end{eqnarray}
 where, $\tilde{y}=(y_1,y_2,\cdots,y_{\frac{D}{2}-1})$. We also need to compute the parity odd letter partition function,
\begin{eqnarray}
i^{Sp(D-2)}_{\tv}(x,\tilde{y})&=& x^2\,\chi^-_{\syng{1,1}}+x^3\, \chi^-_{\syng{2,1}}+x^4\, \chi^-_{\syng{3,1}}\ldots \nonumber
\end{eqnarray}
From the discussion of $Sp(D-2)$ characters in the previous subsections, we immediately notice that the same logical reasoning which enabled to express the infinite sum of $SO(D)$ characters for photons in \cite{Chowdhury:2019kaq}, should also work here. More precisely, we can express $i^{Sp(D-2)}_{\tv}(x,\tilde{y})$ in terms of $i^{Sp(D-2)}_{\ts}(x, \tilde{y})$ in the following manner, 
\begin{eqnarray}
i^{Sp(D-2)}_{\ts}(x,\tilde{y}) ( \,\chi^-_{\syng{1}})&=& \,\chi^-_{\syng{1}} (x+x^2\,\chi^-_{\syng{1}}+x^3\, \chi^-_{\syng{2}}+x^4\, \chi^-_{\syng{3}}+x^5\, \chi^-_{\syng{4}}+\ldots), \nonumber\\
&=& x\,\chi^-_{\syng{1}}+x^2\, \chi^-_{\syng{2}}+x^3\, \chi^-_{\syng{3}}+x^4\, \chi^-_{\syng{4}}+\ldots \nonumber\\
&& \qquad \quad x^2+x^3\,\chi^-_{\syng{1}}+x^4\, \chi^-_{\syng{2}}+x^5\, \chi^-_{\syng{3}}+x^6\, \chi^-_{\syng{4}}+\ldots \nonumber\\
&& \qquad \quad  x^2\,\chi^-_{\syng{1,1}}+x^3\, \chi^-_{\syng{2,1}}+x^4\, \chi^-_{\syng{3,1}}+x^5\, \chi^-_{\syng{4,1}}+x^6\, \chi^-_{\syng{5,1}}+\ldots  \\
\end{eqnarray}

In going from the first line to the second line, we have organised the different tensor products appearing at each power of $x^n$ into three different lines- the first line organises the symmetric tensor, the second line is the anti symmetric trace while the third line organises the antisymmetric traceless tensors. Note the difference from the $SO(D)$ case- the symmetric trace is replaced by the anti-symmetric trace but the parametric structure of the expansion remains the same. We have crucially used the decomposition in \eqref{spsocharacters}. The infinite series in the three different lines can now be expressed in terms of $i^{Sp(D-2)}_{\ts}(x,y)$ with some finite number of subtractions. To summarise, $i^{Sp(D-2)}_{\tv}(x,y)$ is given by, 
\begin{eqnarray}\label{pfph3}
i^{Sp(D-2)}_{\tv}(x,y)&=& i^{Sp(D-2)}_{\ts}(x,y) \chi^-_{\syng{1}} -\left(i^{Sp(D-2)}_{\ts}(x,y)/x -1\right)-x i^{Sp(D-2)}_{\ts}(x,y).
\end{eqnarray}

Using \eqref{pfph2} and \eqref{pfph3}, we can evaluate \eqref{parity2} while \eqref{parity1} is already known from previous results.
\be
I^{D, even}_\tv (x)= \frac{x^8(2+3x^2+2x^4)}{(1-x^4)(1-x^6)}.
\ee 

 As a sanity check, we now perform the large $D$ integral of \eqref{parity2}. Using techniques illustrated in appendix \ref{larged}, we find 

\begin{eqnarray}\label{largedphoton}
I^{D, odd}_\tv (x)= \frac{x^8(2+3x^2+2x^4)}{(1-x^4)(1-x^6)}.
\end{eqnarray}

This implies that there are no parity violating modules in higher dimensions consistent with the observation in \cite{Chowdhury:2019kaq}. Kinematically this is understood as that fact that there are not enough independent vectors to contract with the indices of higher dimensional $\epsilon$ tensors to make a Lorentz singlet.

In lower dimensions we resort to numerical integration outlined in appendix \ref{larged}. In dimensions $D=6$, we obtain 
\begin{equation}
I^{D=6, odd}_\tv (x)=x^8 (2 +3x^2+2x^4-x^8)\denom, \qquad I^{D=6, even}_\tv (x)= x^8 (2 +3x^2+2x^4+x^8)\denom,
\end{equation}
where $\denom= \frac{1}{(1-x^4)(1-x^6)}$. Thus the parity violating module partition function is $I^{D=6,~non-inv}_\tv (x)=x^{16} \denom$ while the parity invariant module partition function is $I^{D=6, inv}_\tv (x)=x^8 (2 +3x^2+2x^4)\denom$. This is consistent with the module partition function generated by \eqref{parity-odd-6d} and \eqref{lagstruct}.

In dimensions $D=4$, the relevant integral gives us 
\begin{equation}
I^{D=4, odd}_\tv (x)=x^8 (1 +x^2+x^4)\denom, \qquad I^{D=4, even}_\tv (x)= x^8 (3 +5x^2+x^4-2x^6)\denom
\end{equation}. 
This implies that in four dimensions the parity violating and parity invariant modules are $I^{D=4,~non-inv}_\tv (x)=x^8 (1+2x^2-x^6)\denom$ and $I^{D=4, inv}_\tv (x)=x^8 (2 +3x^2+x^4- x^6)\denom $ respectively, in accordance with \cite{Chowdhury:2019kaq}. We note that the authors of \cite{Henning:2017fpj} have evaluated this in a different manner as noted in appendix C of their paper. However we find that since we have managed to find closed form expressions for $i^{Sp(D)}_{\tv}(x,y)$, their observation is automatically incorporated since  $i^{Sp(2)}_{\tv}(x,y)=0$.
We also checked numerically that there are no parity violating contributions to the partition function for $D=8$ and $D=10$. We summarize our observations in table \ref{table3}

\begin{table}
	\begin{center}
		\begin{tabular}{|l|l|l|}
			\hline
			Dimension & $I^{D,~inv}_\tv(x)$ & $I^{D,~non-inv}_\tv(x)$ \\
			\hline
			$D\geq  8$ & $x^8(2+3x^2+2x^4)\denom $ & $0$\\
			\hline
			$D=6$ & $x^8(2+3x^2+2x^4)\denom $ & $x^{12}\denom $  \\
			\hline
			$D=4$ & $x^8(2+3x^2+x^4-x^6)\denom$ & $x^8(1+2x^2-x^6)\denom$  \\
			\hline
		\end{tabular}
	\end{center}
\caption{$I^{inv}_\tv(x)$ and $I^{non-inv}_\tv(x)$ partition function for 4 photon S-matrices in even dimensions where $\denom=\frac{1}{(1-x^4)(1-x^6)}$}
\label{table3}
\end{table}

\subsection{Gravitons}
In this subsection we enumerate the partition function corresponding to parity violating four graviton scattering. We review the module counting for gravitons as done in \cite{Chowdhury:2019kaq}. It was established that in $D\geq 8$, the gravitational module is freely generated by one generator at six derivative order, seven generators at eight derivative orders, two generators at ten order in derivatives and finally one generator at twelve derivative order. The local lagrangians generating the modules are polynomials of Riemann tensors and are given by,  
\begin{equation}
\begin{split}
G_{{\bf S},1}&=\,\, \delta_{[a}^{g}\delta_{b}^{h}\delta_{c}^{i}\delta_{d}^{j}\delta_{e}^{k}\delta_{f]}^{l}\,\,R_{ab}^{\phantom {ab}gh} R_{cd}^{\phantom{cd}ij}R_{ef}^{\phantom{ef}kl},\\
G_{{\bf 3},1} &=R_{abpq}R_{baqp}R_{cdrs}R_{dcsr},\\
G_{{\bf 3},2}&=R_{pqrs}R_{pqtu}R_{tuvw}R_{rsvw},\\
G_{{\bf 3},3}&=R_{pqrs}R_{ptru}R_{tvuw}R_{qvsw},\\
G_{{\bf 3},4}&=-R_{pqrs}R_{ptuw}R_{tvws}R_{qvru},\\
G_{{\bf 3},5}&=R_{pqrs}R_{pqtu}R_{rtvw}R_{suvw},\\
G_{{\bf 6}}&=G_{{\bf 3},6}\oplus G_{{\bf 3_A}}=R_{pqrs}R_{pqrt}R_{uvwt}R_{uvws},\\
G_{{\bf 3},7}&=R_{pqab}\partial_a R_{qp \mu\nu} \partial_b R_{rs\nu \alpha} R_{sr \alpha \mu},\\
G_{{\bf 3},8}&=R_{pqab}\partial_a R_{qr \mu\nu} \partial_b R_{rs\nu \alpha}R_{sp\alpha\mu},\\
G_{{\bf S},2}&=R_{abpq} \partial_p \partial_a R_{\mu\nu\beta\gamma}\partial_q\partial_b R_{\nu \alpha\gamma \delta}R_{\alpha\mu\delta\beta}.
\end{split}
\end{equation}

where the generators dual to $G_{{\bf S},1}, G_{{\bf S},2}$ transform in ${\bf 1_S}$, $G_{{\bf 3},1}-G_{{\bf 3},8}$ transform in ${\bf 3}$ while $G_{{\bf 3_A}}$ transforms in ${\bf 3_A}$. A convenient way of describing most of the graviton module is to express it as the product of the photon modules (see \eqref{photon-lag}) since we work to linearised order  and on-shell, 
\be 
R_{abcd} \propto (p_{a}\epsilon_{b}-p_{b}\epsilon_{a})(p_{c}\epsilon_{d}-p_{d}\epsilon_{a})
\propto  F_{ab} F_{cd}. 
\ee
It is not difficult to see that the modules generated by $G_{{\bf 3},1}$ and $G_{{\bf 3},2}$ can be obtained as symmetric product of the photon module $E_{{\bf 3}, 1}$, $G_{{\bf 3},3}$ and $G_{{\bf 3},4}$ can be obtained as symmetric product of the photon module $E_{{\bf 3}, 2}$, $G_{{\bf 3},5}, G_{{\bf 3},6}$ and $G_{{\bf 3_A}}$ is generated by the tensor product  $E_{{\bf 3}, 1} \otimes E_{{\bf 3}, 2}$ while the higher derivative modules $G_{{\bf 3},7}, G_{{\bf 3},7}$ and $G_{{\bf S},2}$ are generated by the tensor product $(E_{{\bf 3}, 1} \oplus E_{{\bf 3}, 2}) \otimes E_{{\bf S}}$ where the photon modules are listed in \eqref{photon-lag}\footnote{The module tensor product $A \otimes B$, denotes that the resulting module elements are given by the set $\{A^{(i)}B^{(i)}\}$ which transform into each other under irreps of $S_3$. For example, $S^2 \left(E_{{\bf 3}, 1}\right)$ generates $\left((E^{(1)}_{{\bf 3}, 1})^2, (E^{(2)}_{{\bf 3}, 1})^2, (E^{(3)}_{{\bf 3}, 1})^2\right)$, which transforms in ${\bf 3}$.}. The six derivative generator dual to $G_{{\bf S},1}$ can not be expressed as product of photon modules (and obviously neither are the descendants). The total contribution to the partition function is 

\be\label{modgravD}
 I^{D,~inv}_{\tt}(x)=x^8(x^{6} Z_{{\bf S}}+ 
5x^{8} Z_{{\bf 3} } + x^{8} Z_{{\bf 6} } + 2x^{10} Z_{{\bf 3} }
+ x^{12} Z_{{\bf S}}).\ee 
A partial classification of the local lagrangians dual to the graviton module was also done in \cite{Fulling:1992vm}. As we move to lower dimensions two things happen- we observe a reduction in the independent generators for the parity-invariant graviton module and there are additional parity violating generators (and corresponding modules) appearing. Similar to the parity-invariant generators, most of the parity violating modules were obtained by tensoring the respective parity violating photon module with the parity invariant counterparts. For brevity we only review the parity violating structures in even dimensions, referring the interested reader to \cite{Chowdhury:2019kaq} for the odd dimensional analysis. 

In $D=8$, it was observed in \cite{Chowdhury:2019kaq}, that the parity even plethystic counted two additional parity violating structures, which are actually total derivative and hence should not contribute to scattering. Thus barring this miscounting, the module partition function should be analogous to the higher dimensional case. The generators are, 
\be
*(F_1\wedge F_2\wedge F_3\wedge F_4){\rm Tr}(F_1 F_2){\rm Tr}(F_3 F_4),\qquad  *(F_1\wedge F_2\wedge F_3\wedge F_4){\rm Tr}(F_1 F_3 F_2 F_4).
\ee

As we can see, as the plethystic counts objects quartic in $F_{\mu\nu}$, it does not recognise that the S-matrix generated by these lagrangians,  $$\varepsilon_{abcdefgh}(k^a_1k^b_2k^c_3k^d_4\epsilon^e_1\epsilon^f_2\epsilon^g_3\epsilon^h_4) \sim 0,$$ by momentum conservation. Descendants of these generators can be explicitly manipulated into total derivatives using bianchi identities, hence the contribution to plethystic (mis)counting is given by, $I^{D=8, non-inv}_{\tt}=2x^{16}$ (Note the absence of $\denom$ which generates the descendants). The corresponding Lagrangians are 
\be
\begin{split}
	\epsilon^{abcdefgh}R_{ab\a\b}R_{cd\b\a}R_{ef\g\d}R_{gh\d\g} \sim *({\rm Tr}(R \wedge R){\rm Tr}(R \wedge R)), \\
	\qquad \epsilon^{abcdefgh}R_{ab\a\b}R_{cd\b\g}R_{ef\g\d}R_{gh\d\a} \sim *({\rm Tr}(R \wedge R \wedge R \wedge R)),
\end{split}
\ee 

where $\wedge$ is taken over eight dimensional space time. 

In $D=6$ the six derivative generator $G_{{\bf S},1}$ is not there and the parity invariant module is given by, $I_{\tt}^{D =6, inv}(x)=x^8( 
5x^8 Z_{{\bf 3} } + x^8 Z_{{\bf 6} } + 2x^{10} Z_{{\bf 3} }
+ x^{12} Z_{{\bf S}})$. We also have three parity violating module generators generated by the parity violating photon module \eqref{parity-odd-6d} and the parity even module \eqref{photon-lag}. Explicitly listed they are,

\begin{eqnarray}
&&H_{{\bf 3_A},1}^{D=6}\equiv  O_{{\bf A}}^{D=6}\otimes E_{{\bf 3},1}, \qquad
H_{{\bf 3_A},2}^{D=6}\equiv  O_{{\bf A}}^{D=6}\otimes E_{{\bf 3},2},\nonumber\\
&&H_{{\bf 3_A},3}^{D=6}=\epsilon^{abcdef}F^1_{ab}(\partial_cF^2_{\mu\nu}\partial_d F^3_{\nu\alpha}F^{4}_{\a \mu})(F^1_{\g\d}F^2_{e\rho}F^3_{f\rho}F^4_{\d\g}),
\end{eqnarray}  
which are respectively generated by the local lagrangians

\begin{eqnarray}
&&\epsilon^{abcdef}R_{\mu\nu\a\b } \partial_\a  R_{\nu\mu ab} \partial_\b R_{\g\d cd} R_{\d\g ef},~~ \epsilon^{abcdef}R_{\mu\nu\a\b } \partial_\a  R_{\nu\g ab} \partial_\b R_{\g\d cd} R_{\d\mu ef} \nonumber\\
&&\epsilon^{abcdef}R_{ab\g\d}\partial_cR_{\mu\nu e\rho}\partial_d R_{\nu\alpha f\rho}R_{\a \mu \d\g}.
\end{eqnarray} 

Note that $H_{{\bf 3_A},3}^{D=6}$ cannot be expressed in terms of products of photon modules. All the modules transform in ${\bf 3_A}$ and contribution to the partition function is given by $I^{ D=6, non-inv}_{\tt}=3x^{18} Z_{\bf{3_A}}$. 

In $D=4$, the parity-invariant generators degenerate further and we just have the generators $G_{{\bf 3},1}, G_{{\bf 3},5}$ and $G_{{\bf S},2}$. Furthermore these do not generate the graviton module freely and there are relations. In summary, the parity invariant module partition function is given by, $I_{\tt}^{D =4, inv}(x)=x^8( 
2x^8 Z_{{\bf 3} } + x^{12} Z_{{\bf S}} -x^{14}Z_{{\bf S}}
-x^{16} Z_{{\bf S}})$. 
The parity violating module is generated by the following generators (see \eqref{parity-odd-4d} and \eqref{photon-lag})
\begin{equation}
H_{\bf 3}^{D=4}=O_{\bf 3}^{D=4}\otimes E_{{\bf 3},1},\qquad H_{\bf S}^{D=4}=O_{\bf S}^{D=4}\otimes E_{{\bf S}},
\end{equation}
which are generated by the lagrangians 
\be 
\epsilon^{abmn}R_{abcd}R_{mndc}R_{opef}R_{pofe},\qquad \epsilon^{abmn}R_{mngh}\partial_a \partial_g R_{cdij}\partial_b \partial_h R_{dejk} R_{ecki}.
\ee 
These modules transform in ${\bf 3}$ and ${\bf 1_S}$ respectively but due to relations in $D=4$ \footnote{See eq (6.47) of \cite{Chowdhury:2019kaq}}, their contribution to the partition function is given by $I^{D=4,~non-inv}_{\tt}=x^8(x^8 Z_{{\bf 3}}(x) + x^{12}Z_{{\bf S}}(x) 
-x^{{14}} Z_{{\bf S}}(x) - x^{{16}}Z_{{\bf S}}(x))$. We evaluate the parity odd plethystic to confirm these observations.

\subsubsection{Parity odd plethystic evaluation}\label{photonpope}
The single letter partition function for gravitons was evaluated in \cite{Chowdhury:2019kaq}.

\begin{eqnarray}\label{grpf1}
i_{\tt}(x,y) &=&  x^4\, \chi_{\syng{2,2}}+x^5\, \chi_{\syng{3,2}}+x^{6}\, \chi_{\syng{4,2}}+x^{7}\, \chi_{\syng{5,2}}+\ldots,\\
&=& x i_{\ts}(x,y) \chi_{\syng{2}} - \left(\frac{i_{\ts}(x,y)}{x}-1-x \chi_{\syng{1}}\right)-\left(x i_{\ts}(x,y) -x^2\right)-\left(i_{\tv}(x,y)-x^2 \chi_{\syng{1,1}}\right)\nonumber\\
&&-x^2 i_{\tv}(x,y) - x^3 i_{\ts}(x,y) ,\nonumber\\
\end{eqnarray} 
where $i_{\ts}(x,y)$ and $i_{\tv}(x,y)$ are defined in \eqref{sodscalar} and \eqref{pfph1} respectively and $\chi_{\syng{2}}, \chi_{\syng{1,1}}$ are respectively the symmetric traceless and anti-symmetric characters of $SO(D)$. For evaluating \eqref{parity2}, we need $i^{SO(D)}_{\tt}(x, \bar{y})$ and $i^{Sp(D-2)}_{\tt}(x, \tilde{y})$, the first of which is straightforward to determine, 
\begin{eqnarray} \label{grpf2}
i^{SO(D)}_{\tt}(x,\bar{y}) &=& x i^{SO(D)}_{\ts}(x,\bar{y}) \chi^{SO(D)}_{\syng{2}}(\bar{y}) - \left(\frac{i^{SO(D)}_{\ts}(x,\bar{y})}{x}-1-x \chi^{SO(D)}_{\syng{1}}(\bar{y})\right)-\left(x i^{SO(D)}_{\ts}(x,\bar{y}) -x^2\right)\nonumber\\
&&-\left(i^{SO(D)}_{\tv}(x,\bar{y})-x^2 \chi^{SO(D)}_{\syng{1,1}}(\bar{y})\right)
-x^2 i^{SO(D)}_{\tv}(x,\bar{y}) - x^3 i^{SO(D)}_{\ts}(x,\bar{y}) ,\nonumber\\
\end{eqnarray}  
We also need the following generating function
\be
i^{Sp(D-2)}_{\tt}(x, \tilde{y}) =  x^4\, \chi^-_{\syng{2,2}}+x^5\, \chi^-_{\syng{3,2}}+x^{6}\, \chi^-_{\syng{4,2}}+x^{7}\, \chi^-_{\syng{5,2}}+\ldots\\
\ee 
This can similarly be obtained from the following sum 
\begin{eqnarray}
i^{Sp(D-2)}_{\ts}(x, \tilde{y}) (x \,\chi^-_{\syng{2}})&=& x\,\chi^-_{\syng{2}} (x+x^2\,\chi^-_{\syng{1}}+x^3\, \chi^-_{\syng{2}}+x^4\, \chi^-_{\syng{3}}+x^5\, \chi^-_{\syng{4}}+\ldots), \nonumber\\
&=&  x^2\, \chi^-_{\syng{2}}+x^3\, \chi^-_{\syng{3}}+x^4\, \chi^-_{\syng{4}}+\ldots \nonumber\\
&& \qquad \quad \,\,\, x^3\,\chi^-_{\syng{1}}+x^4\, \chi^-_{\syng{2}}+x^5\, \chi^-_{\syng{3}}+x^6\, ^-_{\syng{4}}+\ldots \nonumber\\
&& \qquad \quad\,\, \, x^3\, \chi^-_{\syng{2,1}}+x^4\, \chi^-_{\syng{3,1}}+x^5\, \chi^-_{\syng{4,1}}+x^6\, \chi^-_{\syng{5,1}}+\ldots \nonumber\\
&&  \qquad \qquad \qquad \quad \,\, x^4\,\chi^-_{\syng{1,1}}+x^5\, \chi^-_{\syng{2,1}}+x^6\, \chi^-_{\syng{3,1}}+x^7\, ^-_{\syng{4,1}}+\ldots\nonumber\\
&& \qquad \qquad \qquad \quad \,\, x^4\, \chi^-_{\syng{2,2}}+x^5\, \chi^-_{\syng{3,2}}+x^6\, \chi^-_{\syng{4,2}}+x^7\, \chi^-_{\syng{5,2}}+\ldots\nonumber\\
&& \qquad \qquad \qquad \quad \,\, x^4\, +x^5\, \chi^-_{\syng{1}}+x^6\, \chi^-_{\syng{2}}+x^7\, \chi^-_{\syng{3}}+\ldots\\
\end{eqnarray}
Where, similar to the photon case, we have separated the anti-symmetric trace that arises in the tensor product of two $Sp(D)$ tensors according to the order of $x$ they contribute to. This series too can be summed up giving us, 
\begin{equation}\label{grpf3}
\begin{split}
i^{Sp(D-2)}_{\tt}(x, \tilde{y})&=x \,\chi^{Sp(D-2)}_{\syng{2}}(\tilde{y}) i^{Sp(D-2)}_{\ts}(x,\tilde{y})- \left(\frac{i^{Sp(D-2)}_{\ts}(x,\tilde{y})}{x}-1-x \chi^{SP(D-2)}_{\syng{1}}(\tilde{y})\right)\\
&-\left(x i^{Sp(D-2)}_{\ts}(x,\tilde{y})-x^2 \right)-\left(i^{Sp(D-2)}_{\tv}(x,\tilde{y})-x^2 \chi^{Sp(D-2)}_{\syng{1,1}}(\tilde{y})\right)\\
&-x^2 i^{Sp(D-2)}_{\tv}(x,\tilde{y})-x^3 i^{Sp(D-2)}_{\ts}(x,\tilde{y}).
\end{split}
\end{equation}
Using the techniques in appendix \ref{larged} for performing the large $D$ integrals we evaluate the contribution to the parity odd partition function  and recall the parity even contribution worked out before in \cite{Chowdhury:2019kaq},  
\begin{equation}\label{largedgrav}
I^{D, odd}_\tt (x)= I^{D, even}_\tt (x)=\frac{(x^{16} (7 + 10 x^2 + 10 x^4 + 2 x^6 - x^8 + 
x^{10}))}{(1 - x^4 - x^6 + x^{10})}.
\end{equation}
This differs from the module partition function deduced before in \eqref{modgravD} by a factor $-x^{14}+x^{16}$. This miscounting is easily understood due to the generator $G_{{\bf S},1}$, which is cubic in Riemann tensors but it has non-zero contribution from quartic order in $h_{\mu\nu}$ expansion. Since plethystic counts objects quartic in Riemann polynomials, it counts the descendants generated by $G_{{\bf S},1}$ but not the six derivative generator itself. 
$$I^{D, odd}_\tt (x)=I^{D, odd}_\tt (x)=I^{D}_\tt (x)-x^{14}+x^{16}.$$

For low dimensions, we resort to numerical integration and summarise our results in table \ref{table2}. For $D=8$, we obtain the following, 

\be
\begin{split}
	I_\tt^{D=8, even}=x^{16}(9 +10 x^{2}+8 x^{4}-x^{8}+3 x^{10})\denom,~~ 	I_\tt^{D=8, odd}=x^{16}(5 +10 x^{2}+12 x^{4}+4x^6-x^{8}- x^{10})\denom.
\end{split}
\ee 
From table \ref{table2}, we see that this indeed generates plethystic miscounting and it also confirms the conjectured module generators. For $D=6$ and $D=4$, we obtain the following, 

\be
\begin{split}
	I_\tt^{D=6, even}&=x^{16}(6 +9 x^{2}+13 x^{4}+6x^6+3x^{8})\denom,~~ 	I_\tt^{D=6, odd}=x^{16}(6 +9 x^{2}+ 7x^{4}-3x^{8})\denom, \\
	I_\tt^{D=4, even}&=x^{16}(3 +3 x^{2}+5 x^{4}-2x^6-2x^{8})\denom,~~ 	I_\tt^{D=4, odd}=x^{16}(1 + x^{2}+x^{4})\denom.
\end{split}
\ee  
where the even part had been obtained in \cite{Chowdhury:2019kaq}. This is consistent with our module prediction.   
\begin{table}
	\begin{center}
		\begin{tabular}{|l|l|l|}
			\hline
			Dimension & $I^{D,~inv}_\tt(x)$ & $I^{D,~non-inv}_\tt(x)$ \\
			\hline
			$D\geq  10$ & $x^{16} (7 + 10 x^2 + 10 x^4 + 2 x^6 - x^8 + 
			x^{10})\denom $ & $0$\\
			\hline
			$D=8$ & $x^{16} (7 + 10 x^2 + 10 x^4 + 2 x^6 - x^8 + 
			x^{10})\denom $ & $2x^{16}$  \\
			\hline
			$D=6$ & $x^{16} (6 + 9 x^2 + 10 x^4 + 3 x^6 )\denom $ & $x^{16}(3x^4+3x^6+3x^8)\denom $  \\
			\hline
			$D=4$ & $x^{16}(2+2x^2+3x^4-x^6-x^8)\denom$ & $x^{16}(1+x^2+2x^4-x^6-x^8)\denom$\\
			\hline
		\end{tabular}
	\end{center}
	\caption{$I^{inv}_\tt(x)$ and $I^{non-inv}_\tt(x)$ partition function for 4 graviton S-matrices in even dimensions}
	\label{table2}
\end{table}

\subsection{Gluons}\label{gluons}
In this section we generalise the analysis of previous subsections to describe parity violating gluon modules. In \cite{Chowdhury:2020ddc}, the total classification of such module was done and the parity violating contributions conjectured. We will see that our enumeration agrees with their conclusions. As before let us review the classification of \cite{Chowdhury:2020ddc}. The gluon modules can be thought of as tensor product of photon modules and coloured scalar modules- this presents us with two possibilities
\begin{itemize}
	\item Tensor product of modules that are separately quasi invariant 
	 \item Tensor product of modules that are not separately quasi invariant but the tensor product is.  
\end{itemize}.  
In \cite{Chowdhury:2020ddc}, they were labelled as $\CM^{\rm inv}$ and $\CM^{\rm non-inv}$ respectively. The analysis of was done explicitly for $SO(N)$ and $SU(N)$, although the techniques were general. Here for illustration purposes, we consider only gluons charged under adjoint representation of $SO(N)$. 
\subsubsection*{Quasi invariant photon and colour modules}
 The parity odd photon quasi invariant modules were already reviewed in subsection \ref{photons}. In summary, there are no quasi-invariant parity violating photon modules for $D\geq 8$. In $D=6$ there is one six derivative generator transforming in ${\bf 1_A}$ (see eq \eqref{parity-odd-6d}) and in $D=4$ there are two parity violating module generators at four and six derivatives transforming in ${\bf 1_S}$ and ${\bf 3}$ respectively with relations (see eq \eqref{parity-odd-4d}).  
\subsection*{\underline{Colour module}}\label{cmminv}

For $N\geq 9$, it was noted in \cite{Chowdhury:2020ddc} that there are two quasi-invariant color structures ,which were referred to as $\chi_{{\bf 3},1}$ and $\chi_{{\bf 3},2}$.
\bea\label{simple-color}
\chi_{{\bf 3},1}^{(1)}&=&{\rm Tr}(\Phi_1\Phi_2){\rm Tr}(\Phi_3\Phi_4),\qquad
\chi_{{\bf 3},2}^{(1)}={\rm Tr}(\Phi_1\Phi_2 \Phi_3\Phi_4).
\eea
These two structures transform in $\bf{3}$ and do not need explicit $\Z_2 \times \Z_2$ projection. As we move to lower $N$, there are additional structures arising in $N=8, 4$ because of the lower dimensional $\epsilon$ tensor over the colour space. It was observed that such extra quasi-invariant structures were not present for $N=5,6,7$- consequently the number of quasi invariant structures for these values of $N$ was the same as $N\geq9$. 

For $N=8$, we have one additional structure $\chi_{\bf S}^{SO(8)}$ compared to $N\geq 9$, 
\be\label{qiso8}
\chi_{\bf S}^{SO(8)}=\Phi_1\wedge \Phi_2\wedge\Phi_3 \wedge \Phi_4.
\ee
where we use $SO(8)$ epsilon tensor. This generator transforms in $\bf{1}_S$.

For $SO(4)$, we can construct an additional structure using the four dimensional $\epsilon$ tensor (in colour space). This was referred to as $\chi_{\bf 3}^{SO(4)}$  and it transforms in ${\bf 3}$.
\be\label{qiso4sc}
\chi_{\bf 3}^{SO(4),(1)}=\Phi_1\wedge \Phi_2 {\rm Tr}(\Phi_3\Phi_4)|_{\Z_2\times \Z_2}.
\ee
This generator requires explicit  $\Z_2\times \Z_2$ symmetrization. Consequently, it is obvious that this is also one candidate for non-quasi invariant generator- to do so we will explicitly project onto the state with charge $(+--)$ later. 

In summary, for $N\geq 9$, there are two quasi invariant module generator transforming in $\bf{3}$, while for $N=8$ and $N=4$, there is an additional structure transforming in $\bf{1}_S$ and ${\bf 3}$ respectively. Such scalar modules for arbitrary number of $\phi$s have also been classified in \cite{deMelloKoch:2017dgi, deMelloKoch:2018klm}.
 
\subsubsection*{Tensor product: $\mathcal{M}^{inv}$}
The explicit tensor product of the quasi invariant photon module and the colour module generates $\mathcal{M}^{inv}$.  For $D\geq 8$ there are no parity violating gluon modules. 

In $D=6$,  we have one parity violating quasi invariant photon module in $D=6$ \eqref{parity-odd-6d}. For large $N$, we have 2 quasi invariant scalar modules transforming in $\bf{3}$ of $S_3$ \eqref{simple-color}. Using the tensor product rules in appendix \ref{s3review}, we can see that we expect the tensor product to transform in ${\bf 3_A}$. The module generators are given by, 
\begin{equation}\label{d6ngeq9gluon}
\mathcal{G}^{D=6, N\geq 9}_{{\bf 3_A}, 1}= \chi_{{\bf 3},1} \otimes O^{D=6}_{\bf A},\qquad \mathcal{G}^{D=6, N\geq 9}_{{\bf 3_A}, 2}= \chi_{{\bf 3},2} \otimes O^{D=6}_{\bf A}, 
\end{equation}
generated by the following respective lagrangians 
\be 
{\rm Tr} \left( T^e T^f\right) {\rm Tr} \left( T^g T^h\right) ~F^e_{ab}*(\partial^a F^f\wedge \partial^b F^g \wedge F^h),\qquad {\rm Tr}\left( T^e T^f T^g T^h\right) ~F^e_{ab}*(\partial^a F^f\wedge \partial^b F^g \wedge F^h),
\ee 
where $T^\alpha_{\beta\gamma}$ denotes the generator for $SO(N)$ and the trace has been taken over colour space.
The contribution to the partition function for this tensor product is given by $I^{D=6,~non-inv}_{\tg, N\geq 9}=2x^{10}Z_{\bf 3_A}$. The quasi-invariant contribution for $N=5,6,7$ is also the same.

For $N=8$, we have one extra structure \eqref{qiso8}. The extra gluon module generators are given by tensor product 
\begin{equation}\label{d6n8gluon}
\mathcal{G}^{D=6, N=8}_{{\bf 1_A}}= \chi^{SO(8)}_{{\bf S}} \otimes O^{D=6}_{\bf A},
\end{equation}
generated by the following lagrangian 
\be 
\left( T^e \wedge T^f \wedge T^g \wedge T^h\right) ~F^e_{ab}*(\partial^a F^f\wedge \partial^b F^g \wedge F^h),
\ee 
which transforms as ${\bf 1_A}$. The total contribution to the partition function is given by $I^{D=6,~non-inv}_{\tg, N= 8}=2x^{10}Z_{\bf 3_A}+ x^{10}Z_{\bf 1_A}$.

For $N=4$ also we have one extra structure \eqref{qiso4sc}. The extra gluon module generator is given by   
\begin{equation}\label{d6n4gluon}
\mathcal{G}^{D=6, N=4}_{{\bf 3_A}}= \chi^{SO(4)}_{{\bf 3}} \otimes O^{D=6}_{\bf A},
\end{equation}
generated by the following lagrangian 
\be 
\left( T^e \wedge T^f \right){\rm Tr} \left(T^g T^h\right) ~F^e_{ab}*(\partial^a F^f\wedge \partial^b F^g \wedge F^h),
\ee 
which transforms as ${\bf 3_A}$. The total contribution to the partition function is given by $I^{D=6,~non-inv}_{\tg, N=4}=3x^{10}Z_{\bf 3_A}$.

For $D=4$, we have two parity violating quasi-invariant photon modules given by \eqref{parity-odd-4d}. The gluon modules for $N\geq 9$ are generated by 
\begin{equation}\label{d4ngeq9gluon}
\begin{split}
& \mathcal{G}^{D=4, N\geq 9}_{{\bf 3_A}, 1} \oplus \mathcal{G}^{D=4, N\geq 9}_{{\bf 3}, 1}\oplus \mathcal{G}^{D=4, N\geq 9}_{{\bf 3}, 2}=\chi_{{\bf 3},1} \otimes O^{D=4}_{\bf 3},~~ \mathcal{G}^{D=4, N\geq 9}_{{\bf 3}, 5}=\chi_{{\bf 3},1} \otimes O^{D=4}_{\bf S},\\
&\mathcal{G}^{D=4, N\geq 9}_{{\bf 3_A}, 2} \oplus \mathcal{G}^{D=4, N\geq 9}_{{\bf 3}, 3}\oplus \mathcal{G}^{D=4, N\geq 9}_{{\bf 3}, 4}=\chi_{{\bf 3},2} \otimes O^{D=4}_{\bf 3},~~ \mathcal{G}^{D=4, N\geq 9}_{{\bf 3}, 6}=\chi_{{\bf 3},2} \otimes O^{D=4}_{\bf S}.\\
\end{split}
\end{equation}
They are generated by the lagrangians 
\begin{equation}
\begin{split}
\mathcal{G}^{D=4, N\geq 9}_{{\bf 3}, 1}&:{\rm Tr}\left( T^e T^f\right) {\rm Tr} \left( T^g T^h\right) \ve^{\alpha\beta\gamma\delta}F^e_{\alpha\beta}F^f_{\gamma\delta}F_{ab}^g F_{ba}^h,\\
\mathcal{G}^{D=4, N\geq 9}_{{\bf 3}, 2}&:\left({\rm Tr} \left( T^e T^g\right) {\rm Tr} \left( T^h T^f\right) +{\rm Tr} \left( T^e T^h\right) {\rm Tr} \left( T^f T^g\right) \right) \ve^{\alpha\beta\gamma\delta}F^e_{\alpha\beta}F^f_{\gamma\delta}F_{ab}^g F_{ba}^h,\\
\mathcal{G}^{D=4, N\geq 9}_{{\bf 3}_A, 1}&:\left({\rm Tr} \left( T^e T^g\right) {\rm Tr} \left( T^h T^f\right) -{\rm Tr} \left( T^e T^h\right) {\rm Tr} \left( T^f T^g\right) \right) \ve^{\alpha\beta\gamma\delta}F^e_{\alpha\beta}F^f_{\gamma\delta}F_{ab}^g F_{ba}^h,\\
\mathcal{G}^{D=4, N\geq 9}_{{\bf 3}, 3}&:\left({\rm Tr}\left( T^e T^g T^f T^h \right)+{\rm Tr}\left( T^e T^h T^f T^g \right)\right)  \ve^{\alpha\beta\gamma\delta}F^e_{\alpha\beta}F^f_{\gamma\delta}F_{ab}^g F_{ba}^h,\\
\mathcal{G}^{D=4, N\geq 9}_{{\bf 3}, 4}&:{\rm Tr} \left( \{T^e, T^f\}\{T^g, T^h\}\right)  \ve^{\alpha\beta\gamma\delta}F^e_{\alpha\beta}F^f_{\gamma\delta}F_{ab}^g F_{ba}^h,\\
\mathcal{G}^{D=4, N\geq 9}_{{\bf 3_A}, 2}&:{\rm Tr} \left( [T^e, T^f][T^g, T^h]\right)\ve^{\alpha\beta\gamma\delta}F^e_{\alpha\beta}\partial_\gamma F^f_{ab} \partial_\delta F_{bc}^g F_{ca}^h,\\
\mathcal{G}^{D=4, N\geq 9}_{{\bf 3}, 5}&:\left({\rm Tr} T^e T^f\right) {\rm Tr} \left( T^g T^h\right)  \ve^{\alpha\beta\gamma\delta}F^e_{\alpha\beta}\partial_\gamma F^f_{ab} \partial_\delta F_{bc}^g F_{ca}^h,\\
\mathcal{G}^{D=4, N\geq 9}_{{\bf 3}, 6}&:{\rm Tr} \left(T^e T^f T^g T^h\right)  \ve^{\alpha\beta\gamma\delta}F^e_{\alpha\beta}\partial_\gamma F^f_{ab} \partial_\delta F_{bc}^g F_{ca}^h.\\
\end{split}
\end{equation}

The transformation properties of these generators are indicated in the subscript. However from subsection \ref{photons}, we know that for $D=4$, there are relations. Taking the relations into account, the contribution to the module partition function is given by, 
$I^{D=4,~non-inv}_{\tg, N\geq 9}=2x^8(2Z_{\bf 3}+Z_{\bf 3_A})+2x^{10} Z_{{\bf 3}}-2x^{10}(Z_{\bf 3}+Z_{\bf 3_A})$, with the same contribution for $N=5,6,7$. Note that in deriving this, the photon relation in $D=4$ is interpreted as $I^{D=4,~non-inv}_{\tv}(x)=x^8Z_{{\bf 3}}+x^{10}Z_{{\bf S}}-x^{10}Z_{{\bf 2_M}}$ and we use the Clebsch-Gordan products in \eqref{cgoeff1} of appendix \ref{s3review}.

For $N=8$, the extra gluon modules are, 
\begin{equation}\label{d4n8gluon}
\begin{split}
\mathcal{G}^{D=4, N=8}_{{\bf 3}}&=\chi^{SO(8)}_{{\bf S}} \otimes O^{D=4}_{\bf 3},\qquad \mathcal{G}^{D=4, N=8}_{{\bf A}}=\chi^{SO(8)}_{{\bf S}} \otimes O^{D=4}_{\bf S},
\end{split}
\end{equation}
generated by the lagrangians
\begin{equation}
\begin{split} 
\mathcal{G}^{D=4, N=8}_{{\bf 3}}&: \left( T^e \wedge T^f \wedge T^g \wedge T^h\right) \ve^{\alpha\beta\gamma\delta}F^e_{\alpha\beta}F^f_{\gamma\delta}{\rm Tr}F_{ab}^g F_{ba}^h,\\
\mathcal{G}^{D=4, N=8}_{{\bf A}}&: \left( T^e \wedge T^f \wedge T^g \wedge T^h\right)\ve^{\alpha\beta\gamma\delta}F^e_{\alpha\beta}\partial_\gamma F^f_{ab} \partial_\delta F_{bc}^g F_{ca}^h.
\end{split}
\end{equation}
Taking into account the relations, the total contribution to the module partition function is given by, 
$I^{O, D=4}_{\tg, N=8}=x^8(5Z_{\bf 3}+2Z_{\bf 3_A})+x^{10} (2Z_{{\bf 3}}+Z_{{\bf S}})-x^{10}(2Z_{\bf 3}+2Z_{\bf 3_A}+Z_{\bf 2_M})$.

For $N=4$, we have one extra colour quasi invariant structure compared to $N\geq 9$, which generates the following additional gluon modules   
\begin{equation}\label{d4n4gluon}
\begin{split}
\mathcal{G}^{D=4, N=4}_{{\bf 3}, 1}\oplus \mathcal{G}^{D=4, N=4}_{{\bf 3}, 2} \mathcal{G}^{D=4, N=4}_{{\bf 3_A}, 1}&=\chi^{SO(4)}_{{\bf 3}} \otimes O^{D=4}_{\bf 3},\qquad \mathcal{G}^{D=4, N=8}_{{\bf 3}, 2}=\chi^{SO(4)}_{{\bf 3}} \otimes O^{D=4}_{\bf S}.
\end{split}
\end{equation}
They are generated by the lagrangians 
\begin{equation}
\begin{split}
\mathcal{G}^{D=4, N=4}_{{\bf 3}, 1}&:\left( T^e \wedge T^f\right) {\rm Tr} \left( T^g T^h\right) \ve^{\alpha\beta\gamma\delta}F^e_{\alpha\beta}F^f_{\gamma\delta}F_{ab}^g F_{ba}^h\\
\mathcal{G}^{D=4, N=4}_{{\bf 3}, 2}&:\left(\left( T^e \wedge  T^g\right) {\rm Tr} \left( T^h T^f\right) +\left( T^e \wedge T^h\right) {\rm Tr} \left( T^f T^g\right) \right) \ve^{\alpha\beta\gamma\delta}F^e_{\alpha\beta}F^f_{\gamma\delta}{\rm Tr}F_{ab}^g F_{ba}^h\\
\mathcal{G}^{D=4, N=4}_{{\bf 3}_A, 1}&:\left( \left( T^e \wedge T^g\right) {\rm Tr} \left( T^h T^f\right) -\left( T^e \wedge T^h\right)  \left( T^f T^g\right) \right) \ve^{\alpha\beta\gamma\delta}F^e_{\alpha\beta}F^f_{\gamma\delta}F_{ab}^g F_{ba}^h\\
\mathcal{G}^{D=4, N=4}_{{\bf 3}, 2}&:\left( T^e \wedge T^f\right)  \left( T^g T^h\right)  \ve^{\alpha\beta\gamma\delta}F^e_{\alpha\beta}\partial_\gamma F^f_{ab} \partial_\delta F_{bc}^g F_{ca}^h\\
\end{split}
\end{equation}
Taking into the relations of the photon module, we have the following module contribution 
$I^{O, D=4}_{\tg, N=4}=3x^8(2Z_{\bf 3}+Z_{\bf 3_A})+3x^{10} Z_{{\bf 3}}-3x^{10}(Z_{\bf 3}+Z_{\bf 3_A})$.

\subsubsection*{Quasi non-invariant photon and colour modules}
In this subsection we will enumerate the parity violating non-quasi-invariant contribution to the generators of the gluon module. This is given by the tensor product 
$$\mathcal{M}^{non-inv}_{photon,~odd} \tilde{\otimes} \mathcal{M}^{non-inv}_{scalar}$$
Note that the non-quasi-invariant sector also receives contribution from the parity even non-quasi-invariant photon modules but those are not of interest in this present analysis. For asymptotically large dimensions and colour $N$, the non-quasi-invariant generators, be it photon or coloured scalars, can be generated easily- by projecting the generator/colour structure, which don't have automatic $\Z_2 \otimes \Z_2$ symmetry, onto states with charge $(+--)$ rather than $(+++)$. From the analysis in \ref{photons}, we realise that for large dimensions there are no parity violating non-quasi-invariant photon module and hence no contribution to the gluon module from this sector. As we move to lower dimensions, we encounter new parity violating non-quasi-invariant generators which were dealt with case by case in \cite{Chowdhury:2020ddc}. For the colour module, at large $N$, from the discussion in \ref{cmminv}, we see there are no non-quasi-invariant states. For lower $N$, we review below.     

\subsection*{\underline{Photon}}


A natural guess for the non-quasi invariant module generator in 
$D=6$ would be to project the generator \eqref{parity-odd-6d} onto the $(+--)$ state, since the state required explicit $\Z_2\times \Z_2$ symmetrization. However in \cite{Chowdhury:2020ddc}, it was found out that there exists a generator at lower order in derivatives 
\be
\epsilon^{abcdef}F^{ab}_1F^{e\a}_2F^{\a f}_3F^{cd}_4.
\ee
It is not $\Z_2\times \Z_2$ symmetric and it vanishes under $\Z_2\times \Z_2$ symmetrization- recall that we cannot make it non zero by multiplying this with polynomial of mandelstam invariants since mandelstam invariants are automatically $\Z_2\times \Z_2$ symmetric. We can project this generator onto $(+--)$ state to get,
\be\label{parity-odd-6d-qni}
{{\tilde O}}^{D=6,(1)}=\epsilon^{abcdef}F^{ab}_1F^{e\a}_2F^{\a f}_3F^{cd}_4+\epsilon^{abcdef}F^{ab}_2F^{e\a}_1F^{\a f}_4F^{cd}_3.
\ee
It can be verified that the module generated by the anti-symmetric $\Z_2\times \Z_2$ projection of \eqref{parity-odd-6d} onto the $(+--)$ state is a descendant of this module. It transforms as $\bf 3_A$ and we denote this as ${{\tilde O}}^{D=6}_{\bf 3_A}$.

In $D=4$ however, we can use the trick of projecting onto the $(+--)$ state. The parity violating module generators, $O_{\bf 3}^{D=4}$ and $O_{\bf S}^{D=4}$ give rise to the following non-quasi-invariant generators,
\be\label{parity-odd-4d-qni} 
\begin{split}
{\tilde O}_1^{D=4,(1)}&\equiv 2*(F_1\wedge F_2){\rm Tr}(F_3 F_4)|_{(+--)}=4*(F_1\wedge F_2){\rm Tr}(F_3 F_4)-4*(F_3\wedge F_4){\rm Tr}(F_1 F_2),\\
{\tilde O}_2^{D=4,(1)}&\equiv 6\varepsilon_{\mu\nu\rho\sigma}F_1^{\mu\nu} \partial^\rho F_2^{ab}\partial^\sigma F_3^{bc} F_4^{ca}|_{(+--)}=6\big(\varepsilon_{\mu\nu\rho\sigma}F_1^{\mu\nu} \partial^\rho F_2^{ab}\partial^\sigma F_3^{bc} F_4^{ca}+ \varepsilon_{\mu\nu\rho\sigma}F_2^{\mu\nu} \partial^\rho F_1^{ab}\partial^\sigma F_4^{bc} F_3^{ca}\\
&- \varepsilon_{\mu\nu\rho\sigma}F_3^{\mu\nu} \partial^\rho F_4^{ab}\partial^\sigma F_1^{bc} F_2^{ca}-\varepsilon_{\mu\nu\rho\sigma}F_4^{\mu\nu} \partial^\rho F_3^{ab}\partial^\sigma F_2^{bc} F_1^{ca}\big).
\end{split}
\ee
These two generators transform in ${\bf 3}$ and are denoted by ${{\tilde O}}^{D=4}_{{\bf 3},1}$ and ${{\tilde O}}^{D=4}_{{\bf 3},2}$ respectively.

In summary, for $D=6$ there is one parity violating non-quasi-invariant generator at four derivatives transforming in ${\bf 3_A}$ while for $D=4$, there are two generators at four and six derivatives transforming in ${\bf 3}$ and ${\bf 1_S}$ respectively.  
 

\subsection*{\underline{Colour}}

In this section we review the construction of the non quasi -invariant colour modules.  
 
For $N\geq 7$, there exists no non-quasi-invariant structures.

For $N=6$, there exists one non-quasi-invariant structure which is given by 
\be\label{qniso6sc}
\tilde{\chi}^{SO(6),(1)}=\ve_{ijklmn}\Phi_1^{ij}\phi_3^{kl}\Phi_2^{m\alpha} \Phi_4^{n\alpha}|_{(+--)}=\ve_{ijklmn}\Phi_1^{ij}\Phi_3^{kl}\Phi_2^{m\alpha} \Phi_4^{n\alpha}+\ve_{ijklmn}\Phi_2^{ij}\Phi_4^{kl}\Phi_1^{m\alpha} \Phi_3^{n\alpha}.
\ee 
This transforms in a ${\bf 3_A}$ of $S_3$ and we denote it by $\tilde{\chi}^{SO(6)}_{{\bf 3_A}}$.

For $SO(4)$, the color structure $\chi_{\bf 3}^{SO(4)}$, which gave rise to the quasi invariant colour module can also be $\Z_2 \times \Z_2$ anti-symmetrised,
\bea\label{qniso4sc}
{\tilde \chi}^{SO(4),(1)}=\Phi_1\wedge \Phi_2 {\rm Tr}(\Phi_3\Phi_4)|_{(+--)}&=&\Phi_1\wedge \Phi_2 {\rm Tr}(\Phi_3\Phi_4)+\Phi_2\wedge \Phi_1 {\rm Tr}(\Phi_4\Phi_3)\nonumber\\
&-&\Phi_3\wedge \Phi_4 {\rm Tr}(\Phi_1\Phi_2)-\Phi_4\wedge \Phi_3 {\rm Tr}(\Phi_2\Phi_1).
\eea
This structure transforms as ${\bf 3}$ and is denoted by $\tilde{\chi}^{SO(4)}_{{\bf 3}}$. 

In summary, there are no non-quasi-invariant colour modules for $N\geq 7$. For $N=6$ and $N=4$, there is one each, transforming in $\bf{3}_A$ and $\bf{3}$ respectively.

\subsubsection*{Tensor product: $\mathcal{M}^{non-inv}$}
An appropriate tensor product of the non-quasi-invariant modules generate their contribution to the gluon module. There are no non-quasi invariant colour modules for $N\geq 7$.

In $D=6$ and for $N=6$, the colour module and parity violating photon non-quasi-invariant modules are generated by $(+--)$ projections of  $\tilde{\chi}^{SO(6)}_{{\bf 3_A}}$ (\eqref{qniso6sc}) and ${{\tilde O}}^{D=6}_{{\bf 3_A}}$ (\eqref{parity-odd-6d-qni}) respectively.

\begin{equation}\label{d6n6gluonni}
\tilde{\mathcal{G}}^{D=6, N=6}_{{\bf 3}}= \tilde{\chi}^{SO(6)}_{{\bf 3_A}} \tilde{\otimes} \tilde{O}^{D=6}_{{\bf 3_A}}. 
\end{equation}
According to the discussion in section \ref{nqismat}, the tensor product $\tilde{\otimes}$ generates a non-quasi-invariant module transforming in $\bf{3}$ at four derivatives. In equations the action of $\tilde{\otimes}$ generates the following quasi invariant module, 
$$\{|\tilde{\chi}^{SO(6)}_{+--}\rangle | \tilde{O}^{D=6}_{+--}\rangle,~ |\tilde{\chi}^{SO(6)}_{-+-}\rangle |\tilde{O}^{D=6}_{-+-}\rangle,~ |\tilde{\chi}^{SO(6)}_{--+}\rangle |\tilde{O}^{D=6}_{--+}\rangle \},$$
where the subscript denotes the $\Z_2\times\Z_2$ charge of the component module. It is generated by the lagrangian,  
\begin{eqnarray}\label{plagso6adj}
 \ve^{\a\b\g\d\rho\s}T^e_{\a\b}T^f_{\g\d}T^g_{\rho \xi} T^h_{ \sigma \xi} 
~\left(\epsilon^{abcd\xi\zeta}  F^e_{\xi l} F^f_{\zeta l} F^g_{ab} F^h_{cd}\right). \nonumber
\end{eqnarray}
The contribution to the partition function is given by $\tilde{I}^{ D=6, non-inv}_{\tg, N=6}=x^8 Z_{{\bf 3}}$ where $\tilde{I}$ denotes it is the $\CM^{non-inv}$ contribution. 

For $N=4$, the non-quasi-invariant module for colour is given by \eqref{qniso4sc}, which  along with \eqref{parity-odd-6d-qni} generates 
\be\label{d6n4gluonni} 
\tilde{\mathcal{G}}^{D=6, N=4}_{{\bf 3_A}}= \tilde{\chi}^{SO(4)}_{{\bf 3}}  \tilde{\otimes} \tilde{O}^{D=6}_{{\bf 3_A}}. 
\ee  
This transforms in ${\bf 3_A}$ with the following contribution to the partition function $\tilde{I}^{D=6,~non-inv}_{\tg, N=4}=x^8 Z_{{\bf 3_A}}$. The corresponding local lagrangian is given by,
\be
 \ve^{\a\b\g\d}T^e_{\a\b}T^f_{\c\d}T^g_{\rho a} T^h_{ \sigma a} 
~\left(\epsilon^{abcd\xi\zeta}  F^e_{\xi l} F^f_{\zeta l} F^g_{ab} F^h_{cd}\right). \nonumber 
\ee

In $D=4$, we have two non quasi invariant photon generators given by \eqref{parity-odd-4d-qni}. The corresponding contributions by tensoring with $N=6$ non-quasi-invariant colour modules are 

\begin{equation}\label{d4n6gluonni} 
	\begin{split}
\tilde{\mathcal{G}}^{D=4, N=6}_{{\bf 3_A},1}= \tilde{\chi}^{SO(6)}_{{\bf 3_A}} \tilde{\otimes} \tilde{O}^{D=4}_{{\bf 3},1}, \qquad \tilde{\mathcal{G}}^{D=4, N=6}_{{\bf 3_A},2}= \tilde{\chi}^{SO(6)}_{{\bf 3_A}} \tilde{\otimes} \tilde{O}^{D=4}_{{\bf 3},2},
	\end{split}
\end{equation}
with the contribution to the parity violating module partition function given by $\tilde{I}^{D=4,~non-inv}_{\tg, N=6}=(x^8+x^{10}) Z_{{\bf 3_A}}$. 
We do not record the local lagrangian here but it can be easily done analogous to the $D=6$ cases. 

For $N=4$, we have, 
\begin{equation}\label{d4n4gluonni} 
\begin{split}
\tilde{\mathcal{G}}^{D=4, N=4}_{{\bf 3},1}= \tilde{\chi}_{\bf 3}^{SO(4)} \tilde{\otimes} \tilde{O}^{D=4}_{{\bf 3},1}, \qquad \tilde{\mathcal{G}}^{D=4, N=4}_{{\bf 3},2}= \tilde{\chi}^{SO(4)}_{\bf 3} \tilde{\otimes} \tilde{O}^{D=4}_{{\bf 3},2},
\end{split}
\end{equation}
with the contribution to the parity violating module partition function given by $\tilde{I}^{D=4,~non-inv}_{\tg, N=4}=(x^8+x^{10}) Z_{{\bf 3}}$. The total contribution of the parity violating gluon module is a sum of  $\mathcal{M}^{inv}$ and $\mathcal{M}^{non-inv}$ contributions and are summarised in table \ref{pogm}. 
\begin{table}
\begin{center}
	\begin{tabular}{|c | c | c|}
		 \hline
		 & $I^{D=6,~non-inv}_{\tg, N}+\tilde{I}^{D=6,~non-inv}_{\tg, N}$ & $I^{D=4,~non-inv}_{\tg, N}+\tilde{I}^{D=4,~non-inv}_{\tg, N}$ \\ 
		 \hline
		$N\geq 9, N=5,7$ & $2x^{10}Z_{\bf{3}_A}$ & $2x^8(2 Z_{\bf{3}}+ Z_{\bf{3}_A}- x^2Z_{\bf{3}_A})$  \\
		\hline 
		$N=8$ & $2x^{10}Z_{\bf{3}_A}+x^{10} Z_{\bf{1}_A}$& $x^8 \left(-2 \left(x^2-1\right) Z_{\bf{3}_A}+x^2 (Z_{\bf{1}_S}-Z_{\bf{2}_M})+5 Z_{\bf{3}}\right)$\\
		\hline
		$N=6$& $2x^{10}Z_{\bf{3}_A}+x^{8} Z_{\bf{3}}$ & $x^8 \left(4 Z_{\bf{3}}-\left(x^2-3\right) Z_{\bf{3}_A}\right)$\\
		\hline 
		$N=4$ & $3x^{10}Z_{\bf{3}_A}+x^8 Z_{\bf{3}_A}$ & $x^8 \left(\left(x^2+7\right) Z_{\bf{3}}-3 \left(x^2-1\right) Z_{\bf{3}_A}\right)$\\
		\hline  
	\end{tabular} 
\end{center}
\caption{The generating function for parity violating gluon module with the gauge fields charged under the adjoint of $SO(N)$.}\label{pogm}
\end{table} 
\subsubsection{Parity odd plethystic evaluation}
In this section we directly evaluate the gluon plethystic contribution to the multi-particle partition function \eqref{parity1} and \eqref{parity2}. Let us recall the single letter function for gluons charged under the adjoint of $SO(N)$, 
\begin{eqnarray}\label{gluon-single}
i_\tg(x,y,z)&=&{\rm Tr}\,\,x^{\partial} y_i^{L_i}z_\alpha^{H_\alpha}= \chi_{\rm adj}(z) i_\tv(x,y),\nonumber\\
\chi_{\rm adj}(z)&=&\frac{1}{2}\left(\chi_{\syng{1}}(z)^2-\chi_{\syng{1}}(z^2)\right)
\end{eqnarray}
where $\chi_{\syng{1}}(z)$ denotes the character for the vector representation of $SO(N)$.

Thus in comparison with the photon evaluation done in subsection \ref{photonpope}, we have, 
 \begin{equation}\label{parity1gluon}
\begin{split}
I_\tg^{D,~ even}(x) &=\oint  d\mu_{SO(N)}~ \oint  d\mu_{+}~  i_\tg^{(4), +}(x,y,z)/\denom^E_+(x,y).\\
i_\tg^{(4), +}(x,y,z) &=\frac{1}{24}\Big(i_\tv^{SO(D)}(x,y)^4\chi_{\rm adj}(z)^4 +6 i_\tv^{SO(D)}(x,y)^2 i_\tv^{SO(D)}(x^2,y^2)\chi_{\rm adj}(z^2)\chi_{\rm adj}(z)^2\\
&+3i_\tv^{SO(2N)}(x^2,y^2)^2\chi_{\rm adj}(z^2)^2+8i_\tv^{SO(D)}(x,y)i_\tv^{SO(D)}(x^3,y^3)\chi_{\rm adj}(z)\chi_{\rm adj}(z^3)\\
&+6i_\tv^{SO(D)}(x^4,y^4)\chi_{\rm adj}(z^4)\Big),\\
\denom^E_+(x,y)&=\frac{1}{\Pi_{i=1}^{N}\left(1-x y_i\right)\left(1-\frac{x}{y_i} \right)}.
\end{split}
\end{equation}
where the measure $d\mu_{+}$ is the Haar measure corresponding to $SO(D)$ (where $D$ is even) and $d\mu_{SO(N)}$ is the Haar measure corresponding to the colour projection. This was enumerated in \cite{Chowdhury:2020ddc} and we list here in tables \ref{gluonplethd6} and \ref{gluonplethd4} for convenience.
\begin{table}
	\begin{center}
		\begin{tabular}{|l|l|}
			\hline
			$SO(N)$ & $I^{D=6,~ even}_{\tg, N}$  \\
			\hline
			$N\geq 9$ & $2 x^8 (4 + 7 x^2 + 8 x^4 + 4 x^6 + x^8)\denom$\\
			\hline
			$N= 8$ & $x^8 (10 + 17 x^2 + 18 x^4 + 8 x^6 + 3 x^8)\denom$\\
			\hline
			$N=6$ & $x^8 (9 + 15 x^2 + 18 x^4 + 9 x^6 + 3x^8)\denom$\\
			\hline
			$N= 4$ & $x^8 (12 + 23 x^2 + 26 x^4 + 14 x^6 + 3x^8)\denom$\\
			\hline
		\end{tabular}
	\end{center}
	\caption{Partition function $I^{D=6,~ even}_{\tg, N}$ over the space of Lagrangians in $D=6$ involving four $F^a_{\a\b}$'s, where $\denom\equiv 1/((1-x^4)(1-x^6))$. We do not differentiate between $\CM^{\rm non-inv}$ and $\CM^{\rm inv}$ here.}
	\label{gluonplethd6}
\end{table}
\begin{table}
	\begin{center}
		\begin{tabular}{|l|l|}
			\hline
			$SO(N)$ & $I^{D=4,~ even}_{\tg, N}$  \\
			\hline
			$N\geq 9$ & $2 x^8 (6 + 9 x^2 + 7 x^4 + x^6 - 2 x^8)\denom$\\
			\hline
			$N= 8$ & $x^8 (15 + 23 x^2 + 15 x^4 - 4 x^8)\denom$\\
			\hline
			$N=6$ & $x^8 (12 + 19 x^2 + 17 x^4 + 5x^6 - 2 x^8)\denom$\\
			\hline
			$N= 4$ &$ x^8 (19 + 30 x^2 + 24 x^4 + 5x^6 - 6 x^8)\denom$\\
			\hline
		\end{tabular}
	\end{center}
	\caption{Partition function $I^{D=4,~ even}_{\tg, N}$ over the space of Lagrangians in $D=4$ involving four $F^a_{\a\b}$'s, where $\denom\equiv 1/((1-x^4)(1-x^6))$. We do not differentiate between $\CM^{\rm non-inv}$ and $\CM^{\rm inv}$ here.}
	\label{gluonplethd4}
\end{table}
The parity odd contribution is also simple to evaluate since the action of parity is only on the part of the partition function which encodes the space time symmetry, i.e $i_\tv(x,y)$. 
\be\label{glpf1}
i^{Sp(D-2)}_\tv(x,\tilde{y})= \chi_{\rm adj}(z) i^{Sp(D-2)}_\tv(x,\tilde{y}).\nonumber\\
\ee 

Consequently the parity odd only acts on the photon part and the total parity odd plethystic contribution is given by,

\begin{equation}\label{parity2gluon} 
\begin{split}
I_\tg^{D,~ odd}(x) &=\oint  d\mu_{SO(N}~ \oint  d\mu_{-}~  i_\tg^{(4), -}(x,y,z)/\denom^E_-(x,y).\\
i_\tg^{(4), -}(x,y,z) &=\frac{1}{24}\Big(i^{Sp(2N-2)}(x,\tilde{y})^4\chi_{\rm adj}(z)^4 +6 i^{Sp(2N-2)}(x,\tilde{y})^2 i^{SO(2N)}(x^2,\bar{y}^2)\chi_{\rm adj}(z)^2\chi_{\rm adj}(z^2)\\
&+3i^{SO(2N)}(x^2,\bar{y}^2)^2\chi_{\rm adj}(z^2)^2+8i^{Sp(2N-2)}(x,\tilde{y})i^{Sp(2N-2)}(x^3,\tilde{y}^3)\chi_{\rm adj}(z)\chi_{\rm adj}(z^3)\\
&+6i^{SO(2N)}(x^4,\bar{y}^4)\chi_{\rm adj}(z^4)\Big),\\
\denom^E_-(x,y)&=\frac{(1-x^2)}{\Pi_{i=1}^{N-1}\left(1-x y_i\right)\left(1-\frac{x}{y_i} \right)}.
\end{split}
\end{equation}

For large D, we evaluate the integral in \eqref{parity2gluon} using techniques outlined in the appendix \ref{larged} and we confirm that there are indeed no parity violating modules. For lower dimensions, we use numerical integration to compute the parity odd contribution. The results have been summarised in 
tables \ref{gluonplethd6odd} and \ref{gluonplethd4odd}. We find that our results are consistent with the observations of the parity violating module in table \ref{pogm}.   

\begin{table}
	\begin{center}
		\begin{tabular}{|l|l|}
			\hline
			$SO(N)$ & $I^{D=6,~ odd}_{\tg, N}$  \\
			\hline
			$N\geq 9$ & $ 2x^8 (4 + 7 x^2 + 6 x^4 + 2 x^6-2x^8)\denom$\\
			\hline
			$N= 8$ & $x^8 (10 + 17 x^2 + 14 x^4 + 4 x^6-3x^8)\denom$\\
			\hline
			$N=6$ & $x^8 (7 + 13 x^2 + 12 x^4 + 5 x^6-x^8)\denom$\\
			\hline
			$N= 4$ & $x^8 (12 + 21 x^2 + 18 x^4 + 6 x^6 -3x^{8})\denom$\\
			\hline
		\end{tabular}
	\end{center}
	\caption{Partition function $I^{D=6,~ odd}_{\tg, N}$ over the space of Lagrangians in $D=6$ involving four $F^a_{\a\b}$'s, where $\denom\equiv 1/((1-x^4)(1-x^6))$.}
	\label{gluonplethd6odd}
\end{table}
\begin{table}
	\begin{center}
		\begin{tabular}{|l|l|}
			\hline
			$SO(N)$ & $I^{D=4,~ odd}_{\tg, N}$   \\
			\hline
			$N\geq 9$ & $ x^8 (4 + 6 x^2 + 6 x^4 + 2x^6)\denom$\\
			\hline
			$N= 8$ & $x^8 (5 + 7 x^2 + 7 x^4 + 2 x^8)\denom$\\
			\hline
			$N=6$ & $x^8 (4 + 5 x^2 + 5 x^4 + x^6)\denom$\\
			\hline
			$N= 4$ &$ x^8 (5 + 8 x^2 + 8 x^4 + 3x^6)\denom$\\
			\hline
		\end{tabular}
	\end{center}
	\caption{Partition function $I^{D=4,~ odd}_{\tg, N}$ over the space of Lagrangians in $D=4$ involving four $F^a_{\a\b}$'s, where $\denom\equiv 1/((1-x^4)(1-x^6))$.}
	\label{gluonplethd4odd}
\end{table}

\section{Conclusions}
Local  S-matrices  for $2 \rightarrow 2$ scattering can be thought of as module over the ring of polynomials of mandelstam invariants $s, t$ and $u$. The generators for these modules and their descendants are in one to one correspondence with the space of local quartic lagrangians modulo field redefinitions and total derivatives. They can be conveniently expressed in terms of a partition function which encodes the derivative order as well as the $S_3$ properties of these lagrangians. In this note, we have enumerated the parity violating contributions to multi-particle partition function of photons, gravitons and gluons. This encodes the number of independent local contact term like interactions which can contribute to $2\rightarrow 2$ tree-level scattering process which violate parity.  Following \cite{Henning:2017fpj}, we explain how to incorporate the action of parity while evaluating the single letter partition functions in even dimensions, where parity generator does not commute with rotations. We have obtained closed form expressions for the parity odd single letter functions for photons (\eqref{pfph3}), gravitons (\eqref{grpf3}) and gluons \eqref{glpf1} by summing over the $Sp(D-2)$ characters that appear due to the parity action in even dimensions. Taking into account that the normal plethystic exponentiation ( or in the language of \cite{Chowdhury:2019kaq}, Bose exponentiation) has additional terms for the parity odd sector, we evaluated the contributions of these parity odd partition functions to multi particle partition function. The central characters to the story are the parity even and parity odd multi-particle partition function which are summarised in \eqref{parity1} and \eqref{parity2} respectively. The parity invariant and the parity violating module partition function are obtained as a sum and difference of the parity even and parity odd partition function respectively. 

As a sanity check, we obtain the module partition functions at large $D$ which matches previous expectations that for sufficiently large dimensions, we cannot have parity violating interactions contributing to scattering. This can be intuitively understood as the fact that, we do not have enough independent vectors to form Lorentz singlet with an epsilon tensor in high enough dimensions. We exponentiated the relevant $Sp(D)$ Haar measure for this purpose which turns out to be explicitly different from the $SO(D)$ exponentiation. Hence the fact that the parity odd and parity even contributions in large $D$ are the same is a non-trivial consistency check for our formalism. For lower dimensions, we find that the resulting parity odd module partition functions correctly encode the transformation properties of the parity violating lagrangian module generators listed in \cite{Chowdhury:2019kaq, Chowdhury:2020ddc}.

\section*{Acknowledgements}
We thank Kausik Ghosh for discussions during initial stages of the project. The author is supported by a
Kadanoff fellowship at the University of Chicago, and in part by NSF Grant No. PHY2014195.

\appendix 
\section{$S_3$ review}\label{s3review}

The permutation group $S_3$ has three irreps, which can be represented by the following young diagrams 
\be
{\bf 1_S}=\yng(3),\qquad {\bf 2_M}=\yng(2,1),\qquad {\bf 1_A}=\yng(1,1,1).
\ee
where ${\bf 1_S}$ is the totally symmetric one-dimensional irrep, ${\bf 2_M}$ is the two dimensional irrep of mixed symmetry while ${\bf 1_A}$ is the totally antisymmetric one dimensional irrep. The action of $S_3$ on it self gives us the six dimensional representation which can be decomposed into these irreducible representations.
\begin{equation} \label{adjact}
{\bf 6}_{\rm left}= {\bf 1_S}+ 2.{\bf 2_M} + {\bf 1_A} = {\bf 3} + {\bf 3_A}.
\end{equation} 
where ${\bf 3}= {\bf 1_S}\oplus {\bf 2_M}$ and ${\bf 3_A}= {\bf 1_A}\oplus {\bf 2_M}$ respectively are two reducible $S_3$ representations which are used throughout the main text. These are the natural three dimensional representation of $S_3$ with $\Z_2$ even and $\Z_2$ odd charges respectively ($S_3$ is the semi-direct product $\Z_3 \ltimes \Z_2$). Let us consider the action of $S_3$ on the space of polynomials of $s,t,u$ modulo $s+t+u=0$. Its easy to convince oneself that the six dimensional space is spanned by the following generating polynomials with specific $S_3$ transformation properties.

\be \label{partfnapp}
\begin{split}
	&{\bf 1_S}: \qquad (s t u)^m (s^2+t^2+u^2)^n\\
	&{\bf 2_M}_+: \qquad (s t u)^m (s^2+t^2+u^2)^n \{(s + t), (2u^2-s^2-t^2)\},\\
	&{\bf 2_M}_-:\qquad(s t u)^m (s^2+t^2+u^2)^n \{(s- t), (s^2-t^2)\}\\
	&{\bf 1_A}: \qquad (s t u)^m (s^2+t^2+u^2)^n (s^2 t -t^2 s +t^2 u-u^2 t + u^2 s- s^2 u)\\
\end{split}
\ee 
where ${\bf 2_M}_+$ denotes the ${\bf 2_M}$ which has a positive $\Z_2$ charge while ${\bf 2_M}_-$ denotes the one which has a negative $\Z_2$ charge.
As discussed in \cite{Chowdhury:2020ddc, Chowdhury:2019kaq}, the possible partition functions encoding the number of polynomials of $s,t$ transforming in an irrep of $S_3$ at  a particular derivative order is given by,  
\be\label{Z-S3}
Z_{\bf S}=\frac{1}{(1-x^4)(1-x^6)},\quad Z_{\bf M}=\frac{x^2+x^4}{(1-x^4)(1-x^6)},\quad Z_{\bf A}=\frac{x^6}{(1-x^4)(1-x^6)}.
\ee 
corresponding to the three irreducible representations of $S_3$. The partition function for the module transforming in ${\bf 3}$ and ${\bf 3_A}$ therefore becomes

\be \label{3s3apartitionfn}
Z_{\bf 3}=\frac{1+x^2+x^4}{(1-x^4)(1-x^6)}, \qquad Z_{\bf 3_A}=\frac{x^2+x^4+x^6}{(1-x^4)(1-x^6)}.
\ee 

We also write down the Clebsch-Gordan rules which would be used in section \ref{gluons},
\begin{equation}\label{cgoeff1} 
\begin{split}
&{\bf 1_S} \otimes {\bf R}={\bf R},~~~~
{\bf 1_A} \otimes {\bf 1_S}={\bf 1_A},~~~
{\bf 1_A} \otimes {\bf 1_A}={\bf 1_S},\\
&{\bf 1_A} \otimes {\bf 2_M}={\bf 2_M},~~~{\bf 2_M } \otimes {\bf 2_M}= {\bf 1_S} \oplus {\bf 1_A} \oplus {\bf 2_M}.
\end{split}
 \end{equation}
where ${\bf R}$ denotes any of the irreps   ${\bf 1_S}$, ${\bf 1_A}$ or ${\bf 2_M}$.
\section{Details of Haar integral evaluation} \label{larged}
In this section we provide details of the Haar integrals performed in the main text of the paper \ref{po1}. Recall the parity even projection \eqref{parity1} has already been obtained before in literature.The Haar measure for $Sp(D)$, relevant for \eqref{parity2}, is given by 
\be
d\mu_{Sp(D)}=\prod_{i=1}^{\lfloor D/2\rfloor}dy_i \,\tilde{\Delta}(y_i)\,
\ee 
where $\tilde{\Delta}(y_i)$ is the Vandermonde determinant for $Sp(D)$ and is  given by 
\begin{equation}\label{Haarsp}
\tilde{\Delta}(y_i) =\frac{2^{N} \prod_{j=1}^{N} \left( y_j- \frac{1}{y_j}\right)^2 \left(\prod _{j=1}^{N} \left(\prod _{i=1}^{j-1} \left(y_i+\frac{1}{y_i}-y_j-\frac{1}{y_j}\right)\right)\right)^2}{(2\pi i)^N N!\prod_{i=1}^{N}y_i},
\end{equation} 
where $y_i$ are the charges under the cartan subgroup of $Sp(D)$. The integral over $y_i$ is a closed circular contour about $y_i=0$. We need to perform both large $D$ as well as $D=4,6$ evaluation of \eqref{parity2}. In contrast the vandermonde determinant for $SO(D)$ is given by, 
\begin{equation}\label{Haare}
\Delta(y_i) =\frac{2 \left(\prod _{j=1}^{N} \left(\prod _{i=1}^{j-1} \left(y_i+\frac{1}{y_i}-y_j-\frac{1}{y_j}\right)\right)\right)^2}{(2\pi i)^N 2^NN!\prod_{i=1}^{N}y_i}.
\end{equation} 

\subsection*{Large $D$}
In order to perform the large $D= 2N$ Haar integral we set $y_i= e^{i \theta_i}$. We  note that for large $N$, the Haar measure exponentiates, and upto numerical factors it is different from the $SO(2N)$ Haar measure by the following factor, 
\begin{eqnarray}\label{largeNmeasure}
&&\prod_{1\leq j\leq N} \left( \sin \theta_j\right)^2\propto\prod_{1\leq j\leq N} |e^{i \theta_j} - e^{-i \theta_j}|^2 \propto \prod_{1\leq j\leq N} |1 - e^{-2i \theta_j}||1 - e^{2i \theta_j}|,\nonumber\\
&&\propto e^{-\sum_n\frac{\sum^N_{j=1}\left(e^{2n i\theta_j}+e^{-2n i\theta_j}\right)}{n}} \propto e^{-\sum_n\frac{\text{Tr}\mathcal{O}^{2n}}{n}},\nonumber\\
\end{eqnarray}
where $O$ is the orthogonal matrix in diagonal form 
$${\cal O} \sim\left( \begin{matrix}
y_1 & 0 &0 &\cdots &0  \\
0 & \frac{1}{y_1} &0 &\cdots &0\\
0 & 0 & y_2 &\cdots &0 \\
0 & 0 &0 &\frac{1}{y_2}\cdots &0\\
\cdots&\cdots& \cdots& \cdots&\cdots\\
0 & 0 &\cdots&y_N &0\\
0 & 0 &0 &\cdots &\frac{1}{y_{N}}\\
\end{matrix}\right)$$
The Haar measure for the $SO(D)$ in exponential form was already derived in \cite{Chowdhury:2019kaq} and was $\propto e^{-\frac{1}{2n}\sum_n\left(({\rm Tr}{\cal O}^n)^2-{\rm Tr}{\cal O}^{2n}\right)}$. Hence the total exponentiated Haar measure for $Sp(D)$ after accounting for the extra factor, is given by, 
\be\label{largedHaarsp}
\propto e^{-\frac{1}{2n}\sum_n\left(({\rm Tr}{\cal O}^n)^2+{\rm Tr}{\cal O}^{2n}\right)}
\ee 
In the large $N$ limit, $$({\rm Tr}{\cal O}^n)^2 =N^2\rho_n^2,\qquad {\rm Tr}{\cal O}^{2n} = N\rho_{2n}$$. Let us illustrate the large $D$ evaluation for scalars and the photons, gravitons and gluons were done similarly. We can write the integral \eqref{parity2} for scalars as

\begin{equation}\label{parity2scalar} 
\begin{split}
I_\ts^{2N,~ odd}(x) &=\frac{\int {\cal D}\rho_n e^{-\frac{1}{2n}\left(N^2\rho_n^2+ N\rho_{2n}\right)}~  i^{(4), -}(x,y)/\denom^E_-(x,y)}{\int {\cal D}\rho_n e^{-\frac{1}{2n}\left(N^2\rho_n^2+ N\rho_{2n}\right)}}.\\
i_\ts^{(4), -}(x,y) &=\frac{1}{24}\Big(i_\ts^{Sp(2N-2)}(x,\tilde{y})^4 +6 i_\ts^{Sp(2N-2)}(x,\tilde{y})^2 i_\ts^{SO(2N)}(x^2,\bar{y}^2)+3i_\ts^{SO(2N)}(x^2,\bar{y}^2)^2\\
&+8i_\ts^{Sp(2N-2)}(x,\tilde{y})i_\ts^{Sp(2N-2)}(x^3,\tilde{y}^3)+6i_\ts^{SO(2N)}(x^4,\bar{y}^4)\Big),\\
\end{split}
\end{equation}
where we have from \eqref{sodscalar} and \eqref{spdscalar} and the fact that $\tilde{y}^{Sp(2N-2)}= (y_1,y_2,\cdots, y_{N-1})$ and $ \bar{y}^{SO(2N)}= (y_1,y_2,\cdots, y_{N-1},1)$, 
\begin{equation}
\begin{split}
i_\ts^{Sp(2N-2)}(x,\tilde{y})&=x\denom\left(x,\tilde{y}\right), \qquad i_\ts^{SO(2N)}(x,\bar{y})=\frac{x(1-x^2)\denom\left(x,\tilde{y}\right)}{ (1-x)^2},\\
\denom^E_-(x,y)&=(1-x^2)\denom\left(x,y\right), \qquad \denom\left(x,y\right)= \frac{1}{\prod_i \left(1- x y_i \right)\left( 1- \frac{x}{y_i} \right)}.
\end{split}
\end{equation}
 Using the plethystic relation, $\denom(x,y)= e^{\frac{\sum_n x^n \chi_{\syng{1}}(y^n)}{n}}$, we can exponentiate $\denom(x,y)$ as $\denom(x,y)=e^{\sum_{m}\frac{nN\a^n_m\rho_{m}x^{m}}{m}}$ where $\a^n_m$ is non-zero only if $n=0 \mod m$. As an illustrative example and a direct comparision with similar integrals for $SO(D)$, we evaluate the last term of $i_\ts^{(4), -}(x,y)$ in \eqref{parity2scalar},  
\begin{eqnarray} \label{sampplet}
\frac{\int {\cal D}\rho_n e^{-\frac{1}{2n}\left(N^2\rho_n^2+N\rho_{2n}\right)}\denom(x^4,y^4)/\denom (x,y)}{\int {\cal D}\rho_n e^{-\frac{1}{2n}\left(N^2\rho_n^2+ N\rho_{2n}\right)}}	&=& \frac{\int {\cal D}\rho_n e^{-\frac{1}{2n}\left(N^2\rho_n^2- 2N\b^2_n \rho_{n}-8 N \rho _n x^n \alpha^4_n +2N \rho _n x^n \right)}}{\int {\cal D}\rho_n e^{-\frac{1}{2n}\left(N^2\rho_n^2- 2N\b^2_n \rho_{n}\right)}},  \nonumber\\
&=& e^{\left(\sum _{n=1}^{\infty } \frac{x^{2 n} (4 \alpha^4_n-1)^2}{2 n}+\sum _{n=1}^{\infty } \frac{x^n (4 \alpha^4_n-1) \beta^2_n}{n}\right)},\nonumber\\
&=&\frac{1}{-x^6+x^4-x^2+1}.
\end{eqnarray}

where in the second line, we have introduced a fictitious parameter $\b^2_n$ which evaluates to $-1$ only if $n=0 \mod 2$ and also in the third line, we performed the gaussian integrals. Note that the normalisation by the group volume in \eqref{parity2scalar} was crucial to achieve this. Contrast the same integral for the usual parity even Large $D$ (i.e the scalar equivalent of \eqref{parity1}),  done in \cite{Chowdhury:2019kaq} (see Eq H.13). After evaluating every term of \eqref{parity2scalar}, we obtain \eqref{largedscalar}, 
\be 
I_\ts^{D,~ odd}(x) =\frac{x^4}{(1-x^4)(1-x^6)}.
\ee 

This is a non-trivial sanity check of the parity odd plethystic contribution. 

\subsection*{$D\leq 10$}

For $D\leq 10$, we resort to numerical integration. The computation follows very closely to to the method outlined in Appendix H.2 of \cite{Chowdhury:2019kaq} and we encourage the interested reader to look into the details there. In summary, we convert the finite residue integral over $y_i$ to an angular integral by the change of variables $y_i=e^{i\theta_i}$. We expect that the resulting Haar integral will be of the form $f(x)/(1-x^4)(1-x^6)$, where $f(x)$ is a finite polynomial in $x$. Hence we multiply the integrand in \eqref{parity2scalar} by $(1-x^4)(1-x^6)$, expand it as a Taylor series in $x$ and integrate over $\theta_i$ numerically. We find that indeed the integrals yield finite polynomials in $x$ and the results of table \ref{table3}, \ref{table2},  \ref{gluonplethd6odd} and \ref{gluonplethd4odd} have been obtained in this manner.

\end{document}